\documentclass[acmsmall]{acmart}

\usepackage{cleveref}
\usepackage{colortbl}
\usepackage{hhline}     
\usepackage{subcaption}
\usepackage{subfiles}   
\usepackage{xspace}

\microtypesetup{expansion=false}

\AtBeginDocument{%
  }

\newcommand{\edit}[1]{{\textcolor{black}{#1}}}

\newcommand{\eg}{\textit{e.g.}\@\xspace}
\newcommand{\ie}{\textit{i.e.}\@\xspace}

\newcommand{\vs}{\textit{vs.}\@\xspace}

\begin{document}

\setcopyright{acmlicensed}
\acmJournal{PACMHCI}
\acmYear{2025} \acmVolume{9} \acmNumber{7} \acmArticle{CSCW526} \acmMonth{11}\acmDOI{10.1145/3757707}

\title[A Socratic LLM for Asynchronous Deliberation on Perspectivist Data]{Wisdom of the Crowd, Without the Crowd: A Socratic LLM for Asynchronous Deliberation on Perspectivist Data}

\author{Malik Khadar}
\authornote{Both authors contributed equally to this paper.}
\email{khada005@umn.edu}
\orcid{0000-0002-9013-5595}
\affiliation{
  \institution{University of Minnesota}
  \department{Department of Computer Science \& Engineering}
  \city{Minneapolis}
  \country{USA}
}
\author{Daniel Runningen}
\email{runni028@umn.edu}
\orcid{0009-0003-2599-5286}
\affiliation{
  \institution{University of Minnesota}
  \department{Department of Computer Science \& Engineering}
  \city{Minneapolis}
  \country{USA}
}
\authornotemark[1]
\author{Julia Tang}
\email{jyt@umn.edu}
\orcid{0009-0007-4933-2267}
\affiliation{
  \institution{University of Minnesota}
  \department{Department of Computer Science \& Engineering}
  \city{Minneapolis}
  \country{USA}
}

\author{Stevie Chancellor}
\email{steviec@umn.edu}
\orcid{0000-0003-0620-0903}
\affiliation{
  \institution{University of Minnesota}
  \department{Department of Computer Science \& Engineering}
  \city{Minneapolis}
  \country{USA}
  \authornote{Both senior authors contributed equally to this paper.}
}
\authornotemark[2]
\author{Harmanpreet Kaur}
\email{harmank@umn.edu}
\orcid{0009-0009-8239-937X}
\affiliation{
  \institution{University of Minnesota}
  \department{Department of Computer Science \& Engineering}
  \city{Minneapolis}
  \country{USA}
}
\authornotemark[2]

\renewcommand{\shortauthors}{Malik Khadar, Daniel Runningen, Julia Tang, Stevie Chancellor, and Harmanpreet Kaur}

\begin{abstract}
    
Data annotation underpins the success of modern AI, but the aggregation of crowd-collected datasets can harm the preservation of diverse perspectives in data. Difficult and ambiguous tasks cannot easily be collapsed into unitary labels. Prior work has shown that deliberation and discussion improve data quality and preserve diverse perspectives---however, synchronous deliberation through crowdsourcing platforms is time-intensive and costly. In this work, we create a Socratic dialog system using Large Language Models (LLMs) to act as a deliberation partner in place of other crowdworkers. Against a benchmark of synchronous deliberation on two tasks (Sarcasm and Relation detection), our Socratic LLM encouraged participants to consider alternate annotation perspectives, update their labels as needed (with higher confidence), and resulted in higher annotation accuracy (for the Relation task where ground truth is available). Qualitative findings show that our agent's Socratic approach was effective at encouraging reasoned arguments from our participants, and that the intervention was well-received. Our methodology lays the groundwork for building scalable systems that preserve individual perspectives in generating more representative datasets.

\end{abstract}



\received{October 2024}
\received[revised]{April 2025}
\received[accepted]{August 2025}

\maketitle

\section{Introduction}
Data annotation by human crowds is the backbone for the success of many AI systems. Annotators have worked on tasks ranging from image labeling~\cite{rashtchianCollectingImageAnnotations2010} and straightforward content classification~\cite{maSpammerGroupsDetection2024} to more subjective domains, such as content moderation~\cite{ibrohim2019multi,magdy2015quantifying} and toxicity detection~\cite{davidson2017automated}. In many cases, large teams of annotators provide individual labels, and their collective opinions are condensed to a single label~\cite{davidson2017automated}.

\edit{When annotating data for AI, there is a tension between producing datasets that result in accurate, generalizable models, and producing datasets that maintain diverse perspectives.}
Common data curation methods often remove nuance or disagreement in annotations, such as by creating ``ground truth'' datasets using majority voting~\cite{davidson2017automated,alonso2012using,sheshadri2013square}, domain- or expertise-weighted voting~\cite{foody2018increasing,tao2018domain}, crowd consensus via post-hoc clustering~\cite{andre2014crowd,wang2018detecting}, and ``noise removal'' strategies~\cite{khetan2017learning}. Annotations are standardized in a way that does not represent different perspectives. Moreover, those standardized labels risk perpetuating existing systemic harms both directly~\cite{cabitzaPerspectivistTurnGround2023, sandriWhyDonYou2023, sangOriginValueDisagreement2022} and indirectly~\cite{binnsTrainerBotInheritance2017, camboModelPositionalityComputational2022, sambasivanEveryoneWantsModel2021}.\looseness=-1

Treating all differences as noise erases the uncertainty inherent in many kinds of annotation tasks---uncertainty that is relevant if we want models to work for everyone, not just the majority. In response, perspectivist approaches encourage \edit{annotators} to reflect on their opinions, consider their own and others' perspectives, and discuss {\it how} they came to their annotation decisions~\cite{cabitzaPerspectivistTurnGround2023,frenda2024perspectivist}. The term \textit{perspectivism} stems from the belief that data can have multiple ``correct'' interpretations, and that removing voices from a dataset discards nuance and propagates bias~\cite{cabitzaPerspectivistTurnGround2023,camboModelPositionalityComputational2022,kapania2023hunt}. 
Prior work suggests that perspectivist approaches to data annotation and model building \edit{capture the inherent ambiguity of classification tasks more effectively} than their traditional counterparts 
~\cite{davaniDealingDisagreementsLooking2022, gordonDisagreementDeconvolutionBringing2021,leonardelliAgreeingDisagreeAnnotating2021}.
 Therefore, to enable AI models to take more nuanced approaches to problems, we also need datasets that respect nuanced opinions.

Synchronous deliberation between annotators has successfully supported perspectivist approaches for data annotation, but synchronous work has significant resource costs. Design and social science theories argue that deliberation with others richly supports engagement between differing perspectives~\cite{disalvoAdversarialDesign2012,mouffeDeliberativeDemocracyAgonistic1999}.
In practice, research on crowdsourcing has confirmed the value of deliberation in resolving disagreements~\cite{schaekermannResolvableVsIrresolvable2018} and promoting higher-quality annotations through systems like MicroTalk~\cite{drapeau2016microtalk}, Revolt~\cite{changRevoltCollaborativeCrowdsourcing2017}, and Cicero~\cite{chen2019cicero}. However, synchronous deliberation has drawbacks---it demands significantly more time from participants, is challenging to facilitate due to dropout, and is costly. For example, participants in  \citet{schaekermannResolvableVsIrresolvable2018} spent as much as four hours annotating and synchronously deliberating 10 datapoints. Other studies show similar time and cost considerations~\cite{changRevoltCollaborativeCrowdsourcing2017,chenJudgmentSieveReducing2023}.

\edit{In this paper, we introduce a Socratic LLM dialogue system for asynchronous deliberation in data annotation, reaping the perspectivist benefits of synchronous deliberation without the prohibitive costs.} Our system facilitates a Socratic dialogue with annotators, prompting them to reflect on and defend their annotation choices~\cite{abramsReframingSocraticMethod2015,delicSocraticMethodApproach2016,wikiversitySocraticMethods}. Socratic dialogues have been shown to be effective in facilitating meta-cognitive tasks like reflection in educational domains~\cite{boaDevelopmentValidationBlended2018,hoThinkingMoreWisely2023,karguptaInstructNotAssist2024}. We build on the argumentative benefits identified in prior crowdsourcing and deliberation work~\cite{drapeau2016microtalk, chen2019cicero, schaekermannResolvableVsIrresolvable2018} and embody these in an asynchronous workflow using an LLM. As an added benefit, we see this work as a positive use case of LLMs, highlighting a responsible application in a landscape where LLMs are here to stay, yet best practices are still evolving. By limiting the LLM to a Socratic persona, our approach has built-in guardrails to \edit{center annotator perspectives and} avoid known problems of using LLMs (\eg, hallucinations, misrepresentation, and other social biases~\cite{bender2021dangers,li2023dark}).\looseness=-1

To demonstrate the potential of the Socratic LLM for improving perspectivist data annotation, we benchmark the performance of our Socratic LLM against the synchronous deliberation approach of~\citet{schaekermannResolvableVsIrresolvable2018}. We mirror the experimental approach and dataset selection against this work because they provide ample implementation details and publicly released their results for follow-up work~\cite{schaekermannResolvableVsIrresolvable2018}. They test their approach on two datasets well-suited to study the consequences of deliberation on annotation---an inherently ambiguous Sarcasm detection task~\cite{filatova2012irony} and a comparatively objective semantic Relation task~\cite{drapeau2016microtalk}. We compare the two approaches using quantitative metrics such as \edit{the number of annotations that change due to deliberation,} annotator confidence pre- \vs post-deliberation, ground truth accuracy on the Relation task, number and length of dialogues exchanged, and task experience measured with task load index~\cite{hart1988development}. We also report results from qualitatively coding the conversation logs generated from interactions with our system \edit{and the responses to our post-study experiential questions}. \looseness=-1

Our results show that our Socratic LLM performs better than the synchronous deliberation approach of our benchmark on several key metrics. With our approach, participants were more likely to change their labels on the Relation task compared to the benchmark\edit{, with these changes resulting} in more accurate annotations (64.79\% post-deliberation accuracy for us \vs 48.86\% for benchmark); no difference in the Sarcasm task. Not only were participants more accurate, but they were also more confident about their labels after discussion with our Socratic LLM: 28\% of participants switched to higher confidence in our case, compared to no change in confidence \edit{after deliberation in the benchmark}. Finally, our participants were more engaged in their conversations \edit{as} the \edit{amount} and length\edit{s} of their messages \edit{exceeded those in} the benchmark. Our qualitative analysis of conversation logs showed the Socratic LLM played various important roles in deliberation: \edit{as an argument evaluator that questioned annotators' claims, evidence, and warrant; as a negotiator of different class boundaries relevant to the task; as a cognitive support tool that re-articulated prior reasoning; and as a validator that signaled when sufficient levels of deliberation had occurred. }

We highlight the benefits of using a Socratic LLM to assist annotators during the labeling process, proposing the technique as a scalable method for obtaining perspectivist data annotations. Thus, our key contributions are:
\begin{enumerate}
    \item The development of a theoretically grounded Socratic LLM for asynchronous deliberation and insights on its use for a data annotation task.
    \item Quantitative evidence demonstrating the value of our Socratic LLM: improvements in task performance and costs compared to a synchronous deliberation benchmark. 
    \item Qualitative evidence synthesizing the potential roles for Socratic LLMs in annotation \\deliberation workflows.
\end{enumerate}
Finally, we also discuss connections with HCI theories like distributed cognition and boundary objects, articulate the costs and benefits of asymmetric deliberation, consider ethical implications of systems like ours, and present design implications for data annotation tasks and positive LLM roles for future work.\looseness=-1

\section{Related Work}
In this section, we review prior work in three areas: 1) collective annotation, disagreements, and perspectivism; 2) deliberation in annotation; and 3) LLM applications and Socratic dialogue systems. While deliberation and discussion between annotators yield rich datasets that preserve diverse perspectives, coordinating interactions among disjoint annotators introduces immense costs. We posit that deliberation with Socratic LLMs can alleviate the logistical difficulty of taking a perspectivist approach to data annotation and minimizing coordination issues. 

\subsection{Collective Annotation}
Using crowdsourced data annotation to label datasets used by AI systems is a widespread practice~\cite{aroyoTruthLieCrowd2015,destefanoRiseJustTimeWorkforce2015,kitturFutureCrowdWork2013}. However, researchers recognized that there were challenges with the quality crowdsourced labels, due in part to bad data \edit{and} a belief that crowdworkers lacked expertise ~\cite{danielQualityControlCrowdsourcing2018,hallWhatQualityControl2022,irenCostQualityCrowdsourcing2014,nowakHowReliableAre2010,yinPredictingCrowdWork2016}. As a result, crowdsourcing tasks began to gather multiple annotations for each datapoint in a dataset \edit{to improve quality control}~\cite{goyalYourBehaviorSignals2018,liCrowdsourcedDataManagement2016,maFaitCrowdFineGrained2015,quocviethungEvaluationAggregationTechniques2013}.

To handle multiple annotations per datapoint, data curators made a necessary assumption: when decisions were not unanimous, annotations were viewed as \edit{erroneously} ``noisy''~\cite{ibrahimLearningCrowdsourcedNoisy2024,guoLabelCorrectionCrowdsourced2023,zhangImprovingCrowdsourcedLabel2017}. Recently, this assumption has been challenged~\cite{linBiasRepairError2023,sangOriginValueDisagreement2022} as science investigates how to distinguish between bad-faith errors, accidents, uncertainty, and explicit disagreement~\cite{frenayClassificationPresenceLabel2014a,sandriWhyDonYou2023}. Thus, this section reviews the research on resolving annotation {\it errors} versus {\it disagreements.}

\subsubsection{Addressing Errors in Annotations}
Prior work has attempted to identify and resolve noisy data caused by errors~\cite{ibrahimLearningCrowdsourcedNoisy2024,guoLabelCorrectionCrowdsourced2023,zhangImprovingCrowdsourcedLabel2017}. \citet{frenayClassificationPresenceLabel2014a} propose a taxonomy of \edit{noise and methods to reduce it}. For example, researchers have investigated ways to improve the quality of codebooks to mitigate the mistakes caused by task confusion and uncertainty~\cite{braggSproutCrowdPoweredTask2018,k.chaithanyamanamTaskMateMechanismImprove2019,kuoWikibenchCommunityDrivenData2024}. \edit{Some studies} allow the annotators to provide more detailed labels than what practitioners predicted would be necessary (\eg by adding sub-classes for annotation)~\cite{changRevoltCollaborativeCrowdsourcing2017,parkDynamicLabelsSupportingInformed2024}. In \edit{other} cases, the innovation is in communicating uncertainty itself. When dealing with binary choices or sliding scales, \citet{chenGoldilocksConsistentCrowdsourced2021} devised a way for annotators to embed their level of certainty in the label. Finally, researchers have developed algorithms using noisy data to train models to reflect the crowd's collective confidence~\cite{lukasikDoesLabelSmoothing2020}.
Most of these methods, however, only address the problem of ``error'' when it is defined as something to be removed or dealt with.

\subsubsection{Addressing Annotation Disagreement via Perspectivism}
We argue that not all noise is erroneous. Indeed, there are longstanding critiques of attempts to resolve differences in annotation as error~\cite{inel2014crowdtruth, cabitzaPerspectivistTurnGround2023}. Our approach stems from these critiques, and in this section, we overview the ``perspectivist turn'' in data annotation. 

A growing body of work questions the assumptions of a single ground truth always existing in annotation~\cite{aroyoTruthLieCrowd2015,cabitzaPerspectivistTurnGround2023,fleisigPerspectivistParadigmShift2024a,plankProblemHumanLabel2022}, and proposes that, \edit{instead, m}ultiple perspectives are valid when annotating data. The term \textit{perspectivism} is grounded in the belief that data can have various ``correct'' interpretations and that removing voices from a dataset discards nuance that can damage the overall annotation task~\cite{cabitzaPerspectivistTurnGround2023,camboModelPositionalityComputational2022}. For example, existing content moderation models have been shown to disproportionately affect minority groups because of a lack of diversity and missing social context in the training data ~\cite{binnsTrainerBotInheritance2017,registerInitialRemovalLasting2024}. The impact of perspectivism reaches far beyond social media platforms~\cite{chapmanDatadrivenAnalysisInterplay2022,ferrerBiasDiscriminationAI2021}. Perspectivist approaches to annotation and model building enable system designs to be sensitive to social biases~\cite{ferrerBiasDiscriminationAI2021}. There is evidence to suggest that \textit{perspectivist} approaches to model building \textit{perform better} than their traditional counterparts ~\cite{davaniDealingDisagreementsLooking2022, gordonDisagreementDeconvolutionBringing2021,gordonJuryLearningIntegrating2022,leonardelliAgreeingDisagreeAnnotating2021}.


A perspectivist approach can be applied at the annotation level (\eg by soliciting more diverse annotations~\cite{ipeirotis2014quizz,zhuang2015leveraging}), or at the algorithm level. \edit{At the annotation level}, perspectivism acknowledges that data could be interpreted differently and be simultaneously correct depending on the viewpoint. It also means keeping the disagreeing annotations in the resulting datasets instead of resorting to aggregation methods~\cite{gordonDisagreementDeconvolutionBringing2021}. \edit{At} the algorithm level, perspectivism means developing methods for equitably representing these annotation differences in models. Approaches like \citet{wallaceDebiasedLabelAggregation2022} suggest different weighting for models that emphasize nuanced perspectives. Meanwhile, \citet{gordonJuryLearningIntegrating2022} take advantage of diverse annotations to bootstrap a sample population of mock ``jurors'' for deriving a final decision. The one thing these model-building and training approaches have in common is the need for non-aggregated datasets to be made available for research, such as the datasets provided by \citet{schaekermannResolvableVsIrresolvable2018}. Without them, progress on perspectivist approaches will be slow to develop~\cite{cabitzaPerspectivistTurnGround2023,fleisigPerspectivistParadigmShift2024a,plankProblemHumanLabel2022}.

We build on this work by proposing a new method for improving annotation-level consideration of perspectives: one that preserves the voices of all annotators of a dataset. Our method extends prior work on capturing diversity using deliberation, discussed below.

\subsection{Deliberation and its Impact on Annotation}
Deliberation has been proposed to solve sociotechnical problems, where multiple people work together to decide on the outcome of a task or action. Deliberation has been heavily studied in social computing systems because of its relevance to human communication, reasoning, and decision-making (\eg ~\cite{kuoWikibenchCommunityDrivenData2024,zhangDeliberatingAIImproving2023}). Of relevance to our work on perspectivism is deliberation that does not resolve cleanly and that embodies dissensus or disagreement as a reasonable outcome~\cite{mouffeDeliberativeDemocracyAgonistic1999,disalvoAdversarialDesign2012}. Thus, in this section, we explore deliberation and dialogue, when it can lead to disagreement, and what the research says regarding data annotation tasks.


\subsubsection{Deliberation in Annotation Work}\label{sec:most-related-work}
Crowdwork uses discussion between annotators and within teams to resolve disagreements in annotations in a variety of contexts~\cite{changRevoltCollaborativeCrowdsourcing2017,haqbeenPromotingDiscussionAIbased2020,kripleanThisWhatYou2012}.
This is because deliberation improves the quality and explainability of the resulting data~\cite{chingomaDeliberationEvidenceDisclosure2023}, and helps derive a correct answer when the individuals did not know one beforehand~\cite{karadzhovDeliDataDatasetDeliberation2023}.
This type of intervention can be beneficial in simple cases of misunderstanding, but vexing when annotators hold fundamentally different beliefs.
Examining this specifically, \citet{schaekermannResolvableVsIrresolvable2018} studied the impact discussion had on determining if a consensus could be reached on any given annotation item, calling it resolvable \vs irresolvable disagreement. Their results support that reaching consensus is influenced by the nature of the task, but the act of deliberation improves task quality regardless.\looseness=-1

Closest to our work are systems that facilitate dialogue and deliberation for annotation. For example, \citet{drapeau2016microtalk}'s system Microtalk helped  annotators supply additional reasoning for their labels on a relation task. Though MicroTalk did not involve direct discussion, it presented annotators with counter-reasoning in situations where others had already labeled differently. \citet{changRevoltCollaborativeCrowdsourcing2017} expanded on the design of MicroTalk~\cite{drapeau2016microtalk} to allow annotators to better describe the items where they disagreed. In doing so, they helped develop a workflow, Revolt, that generated richer guidelines to describe a dataset. \citet{schaekermannResolvableVsIrresolvable2018} built on MicroTalk~\cite{drapeau2016microtalk} and Revolt~\cite{changRevoltCollaborativeCrowdsourcing2017} to investigate if disagreements between annotators could be reliably resolved through synchronous discussion. 
Their solution involved time-boxed stages to coordinate workers based on their disagreements for the items they labeled. Finally, \citet{chen2019cicero} avoid time-boxed stages by dynamically pairing individuals working on the same task from the start. 

These examples of prior work show the benefits of explicit reasoning and deliberation between annotators, but face challenges regarding worker coordination. We build on this work by eliminating the need for coordinating annotator effort, which allows all annotators to complete their tasks independently. We hope to address the coordination and time constraints by using an LLM as a stand-in discussion partner, benchmarking against the publicly available datasets and results of \citet{schaekermannResolvableVsIrresolvable2018}. 

\subsection{LLMs as Discussion Partners and Socratic Dialogues}
The recent success of LLMs in research, business, and education applications is partly due to their conversational abilities.
Chatbots are now supporting decision-making as mediators for multi-stakeholder applications~\cite{shinChatbotsFacilitatingConsensusBuilding2022}.
Other chatbot research encourages debate with users on video platforms about content to promote media literacy~\cite{tanprasertDebateChatbotsFacilitate2024}.
LLMs are even being used to help guide students through their learning~\cite{al-hossamiCanLanguageModels2024,karguptaInstructNotAssist2024} or perform assessments on their understanding~\cite{hungSocraticMindScalable2024}.
Prior work has also \edit{begun} to explore using conversational LLMs to independently perform and assist with the labeling of datasets~\cite{wu2024usable,maHumanAIDeliberationDesign2024}, but these approaches struggle with the same bias propagation issues as label aggregation methods~\cite{cabitzaPerspectivistTurnGround2023}. We believe in a critical optimist approach to using LLMs: with appropriate guardrails and constrained task responsibilities, we see value in using LLMs for their language generation capabilities. Our approach considers how we can use an LLM to provide a means of reflection with a structured Socratic dialogue.

Many applications of LLMs describe the use of the Socratic method as part of their prompt design considerations --- in debugging code~\cite{al-hossamiCanLanguageModels2024}, teaching math~\cite{dingBoostingLargeLanguage2024}, conducting oral assessment in classrooms~\cite{hungSocraticMindScalable2024}, and evaluating programming assignments~\cite{koutchemeEvaluatingLanguageModels2024}. These classroom-oriented applications come as no surprise, given the studies in educational settings which have long benefited from the use of the Socratic method to promote critical thinking and reflection~\cite{aloniResearchBasedPractices2018,fooDevelopingStudentsCritical2019, walkerActiveLearningStrategies2003}. Because this approach to dialogue is well-established in its use for LLM prompting and meta-cognitive benefits, we find it an appropriate framework for exploring deliberation in dataset annotation. 

We build on prior work by designing a Socratic LLM for an unstructured reasoning task (\ie, annotating subjective and objective data), giving individuals a mechanism to think more critically about their annotations. The following section provides details on the Socratic method and its incorporation into our approach.

\section{Socratic LLM: Design and System}

In this section, we describe our Socratic LLM system design and implementation. This includes design goals and principles, prompt engineering strategies, and finally, an architecture overview of our system implementation.

\subsection{Design Goals}

A major bottleneck of deliberation is the need for two or more people to actively and synchronously engage in discussion~\cite{changRevoltCollaborativeCrowdsourcing2017,chen2019cicero}. Additionally, designers need to ensure the mode of deliberation is appropriate for the task at hand, such as debates~\cite{walton1998new} or panel~\cite{williams2014harkness} formats.

We developed a {\bf Socratic LLM} to act as an inquisitor from the Socratic Method to address these challenges. Our goal was to provide users with a conversational partner that provokes moments of reflection while limiting auxiliary input. First, our Socratic LLM helps lower logistical constraints by supporting asynchronous deliberation with an individual annotator. Second, recent work in programming education (\eg,~\cite{karguptaInstructNotAssist2024,al-hossamiCanLanguageModels2024,koutchemeEvaluatingLanguageModels2024}) and mathematics~\cite{dingBoostingLargeLanguage2024} has demonstrated the efficacy of LLMs to act as a Socratic \textit{instructor and inquisitor} to support meta-cognitive skills. We conceptualize this process similarly to \citet{hungSocraticMindScalable2024}'s work using a Socratic agent in educational settings. Our application is similar to education domains by engaging in higher-order cognitive processes to facilitate an outcome instead of serving as a question-answering system~\cite{abramsReframingSocraticMethod2015,aloniResearchBasedPractices2018,boaDevelopmentValidationBlended2018,delicSocraticMethodApproach2016,fooDevelopingStudentsCritical2019}.

\begin{figure}
    \centering
    \includegraphics[width=\linewidth]{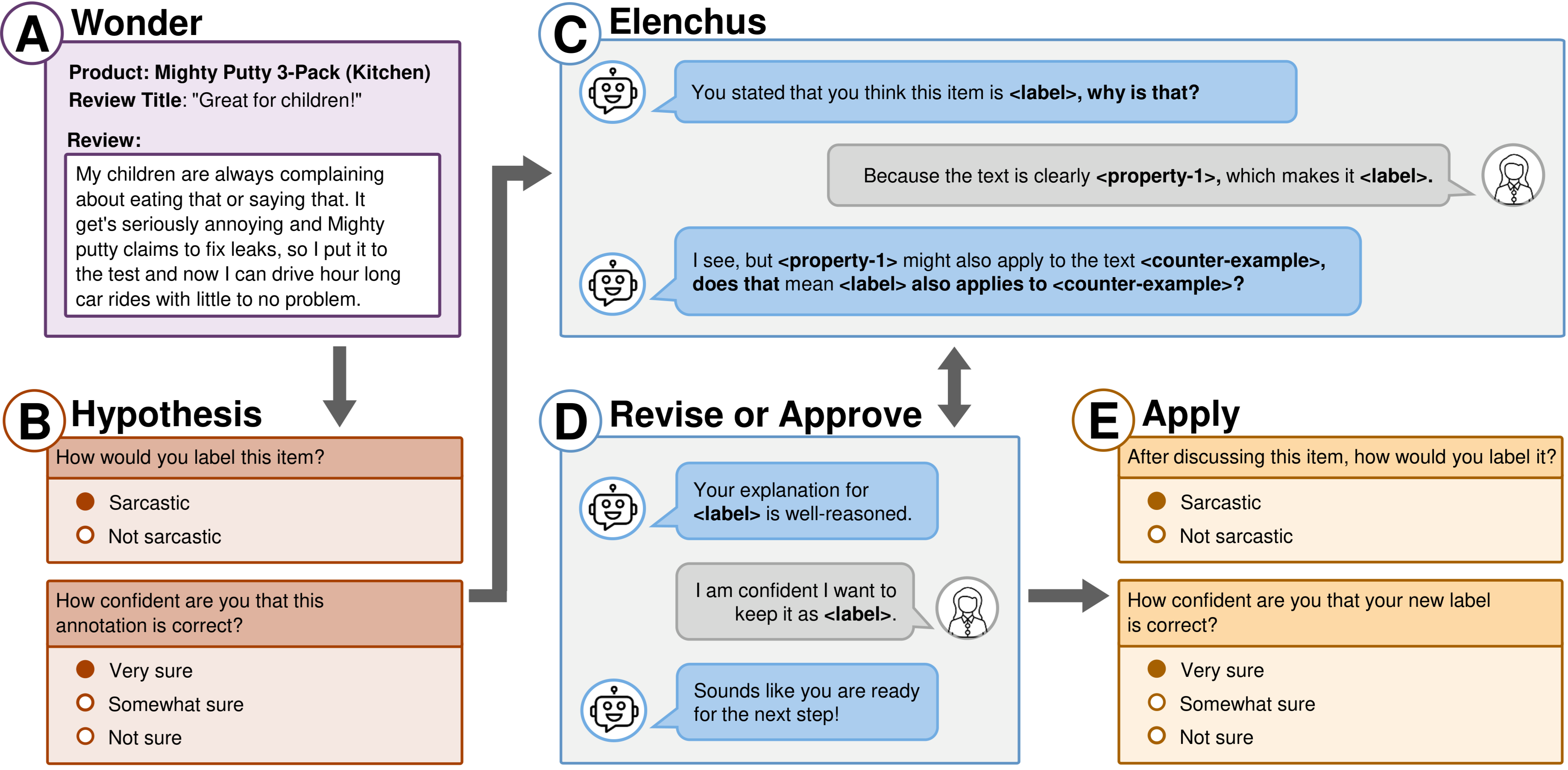}
    
    \caption{\textbf{Socratic Method in practice.} Participants \edit{progress} through the phases of the Socratic Method as they complete annotation tasks. (A) Wonder, the participant considers the datapoint provided. (B) Hypothesis, participants generate a hypothesis based on the data and respond to questions to record their annotation. (C) Elenchus, engagement with the Socratic LLM encourages participants to reflect on their hypothesis and explain their position. (D) \edit{Revise or Approve}, \edit{if their hypothesis does not withstand scrutiny, then participants revise it and return to the discussion style of Elenchus. Otherwise, their hypothesis is collectively approved.} (E) \edit{Apply, participants record their final annotation based on their approved hypothesis.}}
    \label{fig:soratic-method}
    \Description{Diagram of the general flow of the Socratic Method through the perspective of an example interface.}
\end{figure}

Our Socratic LLM facilitated the latter half of the Socratic Method~\cite{wikiversitySocraticMethods}. The crowdworker independently 
completed the two initial stages---\textit{wonder} and \textit{hypothesis}---before engaging with the Socratic LLM in {\it elenchus}: ``a form of cooperative argumentative dialogue between individuals, based on asking and answering questions to stimulate critical thinking and to draw out ideas and underlying presumptions''~\cite{wikiversitySocraticMethods}. Then, we asked participants to engage in {\it approving or revising of hypothesis:} if the hypothesis fails to stand up to scrutiny, the discussion returns to elenchus. Once the hypothesis is approved, participants can move to the final stage of \textit{application}, where the new hypothesis can be used to revise one's understanding or behavior. Figure~\ref{fig:soratic-method} shows how the Socratic Method maps onto our annotation process.

\subsection{Prompt Engineering}\label{sec:prompt-engineering}
To ensure a successful dialogue, we designed three components \edit{in} our prompt: instructions for performing the elenchus, guidelines on a ``Socratic Temperament''~\cite{wikiversitySocraticMethods}, and general guardrails to keep responses manageable. The full, human-readable prompt is provided in Appendix~\ref{app:prompt}.

{\bf Socratic Processes.} First, prompt engineering and development focused on the process of elenchus, the cooperative dialogue that interrogates the participants' hypothesis. The goal of the Socratic dialogue system is to help revise and approve the hypothesis before the user applies the result of their hypothesis during re-annotation. To do this, five steps were provided in the system prompt to the LLM, mapping to prior work on Socratic systems (\eg~\cite{hungSocraticMindScalable2024,wikiversitySocraticMethods}). The first step expects the participant to assert a claim based on their label. This step is followed by the dialogue system interrogating the claim, either by asking clarifying questions or prompting further reflection to illustrate potential gaps in logic. The participant is then expected to engage with the dialogue system's line of inquiry. If inconsistencies are identified, the Socratic LLM should ask the participant to update their claim and restart with the new claim. If the reasoning is sound, the participant should be encouraged to affirm their hypothesis using the annotation interface and move forward.

{\bf Temperament.} Second, we wanted the Socratic LLM to maintain a reasonable temperament that supported inquisitive engagement. To \edit{that end}, our prompt outlined four primary traits that emerged through pilot testing: humility, respect, joy in the dialogue, and prioritizing mutual understanding. Temperament is key for dialogue systems like ours~\cite{chaves2021should}, and we were mindful that the process of elenchus can come across as overly and unintentionally forceful or manipulative~\cite{wikiversitySocraticMethods}. The Socratic LLM needed to remain impartial to ensure that the dialogue was centered on the participant's developing hypothesis rather than biases expressed by the Socratic LLM.

{\bf Guardrails.} Finally, we added guardrails and rules to the prompt to ensure the responses were manageable for participants in a crowdworking task. We instructed the LLM not to use outside knowledge; we did not want to bias annotators with ground truth data available through the LLM's internal datasets or outside resources it had scraped (\eg, Wikipedia data may confirm a Relation annotation). Through extensive pilot testing, we also identified structural rules necessary to keep the Socratic LLM's responses dialogic rather than prescriptive, ensure good conversation practices, and avoid unwanted behaviors. For example, we included low-level instructions about the interaction format, like limiting the responses to two or three sentences, as LLM responses can otherwise be quite long~\cite{nayab2024concise}. Other rules helped constrain the conversation to promote the turn-taking intentions of the Socratic process and prevent the Socratic LLM from getting off-task.

\subsection{Architecture}

{\bf LLM.} Our LLM was built with Anthropic's Claude 3 Haiku model\footnote{\url{https://www.anthropic.com/news/claude-3-haiku}}. The prompt underwent six major iterations before pilot testing showed behavior that sufficiently reflected the Socratic process.

\noindent{\bf System Architecture.} Our system was built on the Express framework\footnote{\url{https://expressjs.com}} for the back-end and VueJS framework\footnote{\url{https://vuejs.org}} for the user interface\edit{, with styling implemented via Bootstrap libraries\footnote{\url{https://getbootstrap.com}}}. Data storage was handled by an instance of MySQL server\footnote{\url{https://www.mysql.com}} and retrieved through RESTful API calls with our back-end. Our LLM was integrated as a service in Express using Anthropic's API interface.\footnote{\url{https://www.anthropic.com/api}}\looseness=-1

\section{Experimental Design}
We based our experimental design on our intended benchmark, \citet{schaekermannResolvableVsIrresolvable2018}, which examined how synchronous deliberation helps crowdworkers resolve disagreements on data annotation tasks. Though we discuss several studies from prior work that are related to our goals (see Section~\ref{sec:most-related-work}), \citet{schaekermannResolvableVsIrresolvable2018}'s work uniquely affords benchmarking given their publicly-available data and results. We designed a study to evaluate the impacts of our Socratic LLM on annotation outcomes, mirroring the experimental design and artifacts used by our benchmark. Our research was reviewed and approved by \edit{the University of Minnesota’s Institutional Review Board (IRB)}.\looseness=-1

\subsection{Benchmark Datasets}
We used the \edit{following} two datasets from \citet{schaekermannResolvableVsIrresolvable2018}, each containing 40 items which we refer to as ``datapoints.''

{\bf Sarcasm Dataset.} The first dataset asked annotators to identify whether an Amazon product review was sarcastic.
Sarcasm detection is considered a subjective task because there is no universal definition of sarcasm, making it well-suited for deliberation. For this task, \citet{schaekermannResolvableVsIrresolvable2018} built their dataset from the sarcasm detection of \citet{filatova2012irony}, filtering their dataset to identify the labels with the highest inter-rater disagreement. An example from this dataset is included in our system screenshots under Appendix~\ref{app:screenshots}.

{\bf Relation Dataset.} The second dataset asked annotators to identify whether a particular relation between two objects in a sentence was present. Depending on the datapoint, the relation would either be ``lived in'' or ``died in.'' This is a common benchmark in deliberation tasks~\cite{drapeau2016microtalk,chen2019cicero} and is considered objective because there are official labeling guidelines on {\it LivedIn} and {\it DiedIn} relations published by the Linguistic Data Consortium.\footnote{\url{https://tac.nist.gov/2017/KBP/index.html}} However, individual datapoints can be ambiguous \edit{and thus lack a ground truth label}. Our specific dataset was taken from MicroTalk~\cite{drapeau2016microtalk} and modified by our benchmark~\cite{schaekermannResolvableVsIrresolvable2018} to clarify the task instructions. One example from this dataset asked if the relationship ``Litvinenko `lived in' London'' was expressed in the sentence ``Litvinenko died of radioactive poisoning, from Polonium 210, in his home in London.''

\subsection{Study Task}

\begin{figure}
    \centering
    \includegraphics[width=\linewidth]{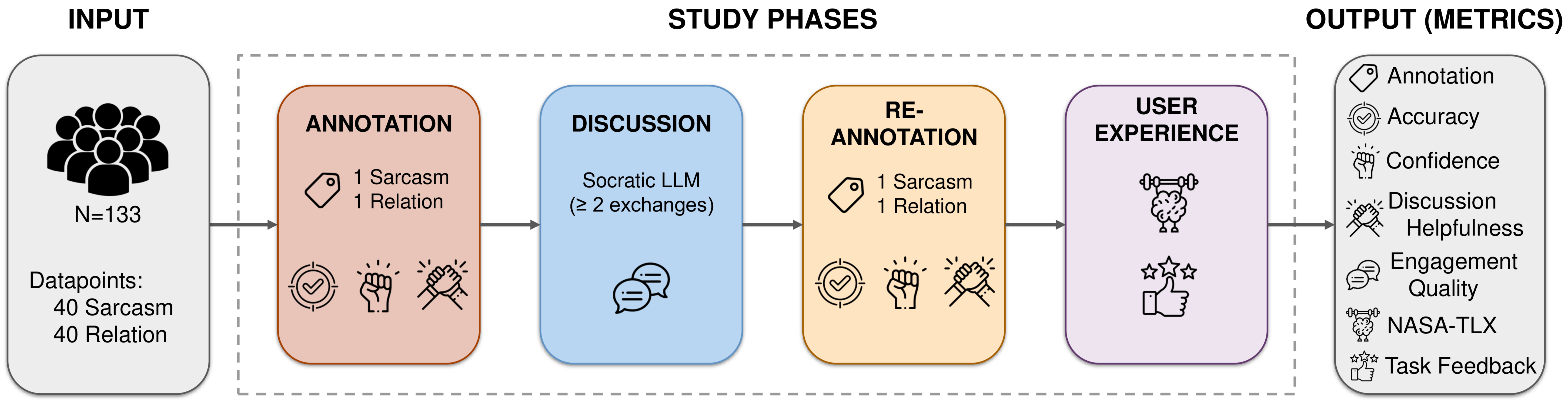}
    \caption{Workflow diagram of the Socratic LLM-assisted annotation process. Crowdworkers (N=133) first performed initial labeling tasks for Sarcasm and Relation detection, followed by an AI-driven discussion phase using Socratic questioning. \edit{The crowdworkers} then reconsidered their annotations and confidence levels before completing NASA-TLX and experience assessments.}
    \label{fig:process_socratic}
    \Description{Diagram showing the different phases of the annotation process and their outputs.}
\end{figure}

Our study was split into four phases: (1) an annotation phase where participants labeled datapoints from the datasets; (2) an asynchronous discussion phase with our Socratic LLM, where participants reasoned about their annotations; (3) a re-annotation phase where they had the opportunity to reconsider their labels; and (4) a user experience survey. Figure~\ref{fig:process_socratic} includes an overview of these phases. The exact wording and setup of the questions can be found in Appendix~\ref{app:all-questions}, and example screenshots can be found in Appendix~\ref{app:screenshots}. 

\subsubsection{Annotation Phase}

First, participants were asked to annotate two randomly-selected datapoints, one each from the {\bf Sarcasm} and {\bf Relation} datasets.
Only one datapoint was displayed at a time. The task included the same context and instructions as our benchmark for consistency.
We randomly sampled from each dataset until at least three participants had labeled each datapoint. After annotating each datapoint, participants rated: (1) their confidence that their annotation was correct (options: ``Very Sure'', ``Somewhat Sure'', and ``Not Sure''); and (2) whether having a discussion would \edit{reduce} their uncertainty about their label (options: ``Yes'' or ``No'').

We also placed attention check questions among those listed above, one for each annotation (see Appendix~\ref{attentionQs} for the two attention checks). If participants failed both attention \edit{checks}, they were removed from the study, and their data was excluded from analysis. After annotating, participants were asked to confirm that they wished to proceed and warned that they would not be able to return to this phase.

\subsubsection{Discussion Phase with Socratic LLM}
Participants then entered the Discussion Phase, engaging with the Socratic LLM. In addition to reviewing their datapoint and the annotation instructions, participants had a chat interface to converse with the dialogue system (see Figure~\ref{fig:re-annotation} in Appendix~\ref{app:screenshots}). They were instructed to discuss their reasoning for each annotation. Thus, participants engaged in two distinct chat sessions with the Socratic LLM, one for each datapoint.\looseness=-1

Participants could submit as many messages as desired to our Socratic LLM but were required to respond at least twice to each discussion before proceeding. This mirrored the minimum participation required from collaborative discussions in our benchmark.\looseness=-1

\subsubsection{Re-annotation Phase}
After chatting with our Socratic LLM, participants were allowed to re-annotate each datapoint, with clear communication that changing their labels was not required. This phase was critical in capturing the impact of our asynchronous discussion and included questions about both the (re-)annotation and the influence of the conversation (see Appendix~\ref{reAnnotationQs} for exact questions). After the re-annotation phase of the first datapoint, we shared a short 12-second GIF\footnote{\url{https://media1.tenor.com/m/2EXMLYXzJAYAAAAd/dog-stand-up-dog-wake-up.gif}} as a transition to give participants a \edit{short} break before discussing the second datapoint.

\subsubsection{User Experience}
\label{methods:post-questions}
Finally, participants were directed to answer experiential questions \edit{in} a Qualtrics survey.\footnote{\url{https://www.qualtrics.com}}
These questions were focused on prior deliberation experience and task difficulty to help lay the foundations of comparison for future work.

We measured task difficulty using five of the NASA Task Load Index (NASA-TLX) questions~\cite{hart1988development}. We omitted the sixth question on the physical demand of the task, following guidance from prior work where NASA-TLX has been used to compute task load in computational settings~\cite{cai2016chain,iqbal2005investigating,kaur2018creating}. Additionally, we asked several Likert-scale questions to measure current and prior annotation experience, and \edit{to determine the perceived helpfulness of the Socratic LLM}; yes/no questions about opinions on and experiences with deliberation for annotation; and open-ended questions for feedback about the experience (see Appendix~\ref{expQs} for exact questions).

\subsection{Participants, Consent, and Procedure}

We recruited 133 participants on Prolific to participate in our experiment.\footnote{\url{https://www.prolific.com}} Prolific is an online research platform where participants can be recruited for crowdsourcing tasks. The study was released in batches, with approximately 30 slots in each batch to allow for manual verification of work and \edit{for} observation of system functionality.

We used Prolific's built-in screening mechanisms to filter participants by location (restricted to the United States) and English language fluency. Our participants ranged in age from 18 to 70 (mean = 37.23, sd = 10.87). 
56.39\% of participants identified as female through the Prolific platform, 42.86\% as male, and 0.75\% selected ``Prefer not to say''. 
\edit{61.65\% of the sample identified themselves as White, 21.05\% as Black, 7.52\% as Asian, 7.52\% as Mixed, and 2.26\% as Other.}\footnote{Prolific's platform provides only the options listed in this section to describe a user's legal sex and ethnic group. We recognize that these categories may not fully represent the identities of our participants and note this as a limitation.}

To receive compensation of \$5 USD, participants had to meet our selection criteria, opt into participation on Prolific, confirm consent via a linked Qualtrics survey, and complete all study tasks. We disqualified participants for: 1) failing our attention checks; 2) when there were clear signs of misconduct, such as willfully ignoring task instructions; 3) providing responses generated from outside sources like ChatGPT; or 4) when their task completion speed for their provided responses was much too high to have been typed by hand. Participants could opt out of this study at any time until payment was submitted. 

\edit{Based on pilot testing, we} estimated this task would take \edit{approximately} 20 minutes to complete. On average, participants spent 20.5 minutes (sd = 11.7) on the study, with 15.4 minutes (sd = 8.3) spent on the annotation and deliberation sections. \edit{We targeted our compensation at \$15 USD/hr.}

\subsection{Analysis and Baseline Comparisons}
Our analysis involved quantitative evaluations of the participants' annotations and qualitative assessments of their messages with the LLM and their experiences with the study.

The quantitative data we collected was directly compared to data and evaluations provided by~\citet{schaekermannResolvableVsIrresolvable2018}. Using this prior work and their metrics as a benchmark, we analyzed:
\begin{enumerate}
    \item \textbf{Changes in Annotations. }The frequency of annotation changes after discussions. This includes changes in annotations generated by an individual before \vs after discussion, and across all annotations for a datapoint.
    \item \textbf{Accuracy of Relation Data Annotations. }The accuracy of final annotations for datapoints where ground truth was available for the Relation dataset.
    \item \textbf{User Confidence. }Participants' confidence in their annotations before \vs after discussion.
    \item \textbf{Engagement Quality. }The length and number of messages sent by individual participants.
\end{enumerate}

\noindent The benchmark did not always have a perfectly matched population to ours to serve as a baseline due to the constraints associated with synchronous deliberation and their time-boxed staging approach. We note two important cases of this. \textit{First}, for metrics 1--3, we identify the subset of annotations from~\citet{schaekermannResolvableVsIrresolvable2018} where participants completed all task assignments and thus would be a fair comparison to our case. This included some annotations where there was not enough disagreement to require discussion. 
\textit{Second}, for the Engagement Quality metric, note that the benchmark's and our analysis is contingent upon deliberation. Therefore, we only use data from \edit{items that were deliberated on} in both our study and the benchmark to ensure a fair comparison. Although we could leverage other engagement quality metrics (\eg, time) for comparison, these would be unfair to the benchmark's time-boxed staging approach and always show results in our favor (details in Section~\ref{sec:engagement-quality}). Our metrics reflect what was collected and measured by ~\citet{schaekermannResolvableVsIrresolvable2018}, excluding metrics yielding unfair or unreasonable comparisons.

For qualitative coding, we conducted inductive thematic analysis~\cite{braun2006using}. The \edit{first} author open-coded the LLM chat logs, and the third author open-coded the open-text study experience responses. A review of all open codes and axial coding was jointly conducted between the first and second authors. Finally, all authors discussed the axial codes and collaboratively extracted themes.

\section{Findings}

Our results show the Socratic LLM promoted perspectivist thinking in the annotation process and outperformed the synchronous deliberation benchmark along several key metrics. Comparing to \citet{schaekermannResolvableVsIrresolvable2018}'s work as a benchmark, our main findings are:
\begin{enumerate}
    \item Participants' switched annotations significantly more \edit{often} after discussion with the Socratic LLM for Relation tasks.
    \item Importantly, these changes to the Relation annotations from interacting with the Socratic LLM improved annotation accuracy post-deliberation.
    \item Overall accuracy of our final Relation dataset labels was higher compared to the benchmark.
    \item Confidence in annotations increased after engagement with the Socratic LLM.
    \item The quality of our participants' engagement was higher on average than the benchmark.
    \item Our qualitative analysis of conversations with the Socratic LLM identified different LLM roles that embodied the Socratic processes and facilitated better reasoning.
\end{enumerate}

\subsection{Changes in Annotations}

We define a ``flip'' as an annotation that changed pre- \vs post-deliberation. When analyzing flips, we focus on the changes to the binary labels (\eg, expressed/not expressed and sarcasm/not sarcasm) and omit post-deliberation labels of ``Not Sure'' in both the benchmark and our data.\footnote{In our study, there were 45 annotation-level flips. Of these, 6 were changed to ``Not Sure'' (or 2.25\% of the total), evenly split between the two datasets with no overlap on a particular datapoint or participant. In the benchmark, there were 43 changes to ``Irresolvable'' for the Sarcasm dataset (3.02\% of 1424 total annotations) and 61 for the Relation dataset (3.22\% of 1896 total annotations).}\looseness=-1

\subsubsection{Annotation-Level Flips}

\begin{figure}[t]
    \centering
    \includegraphics[width=1\linewidth]{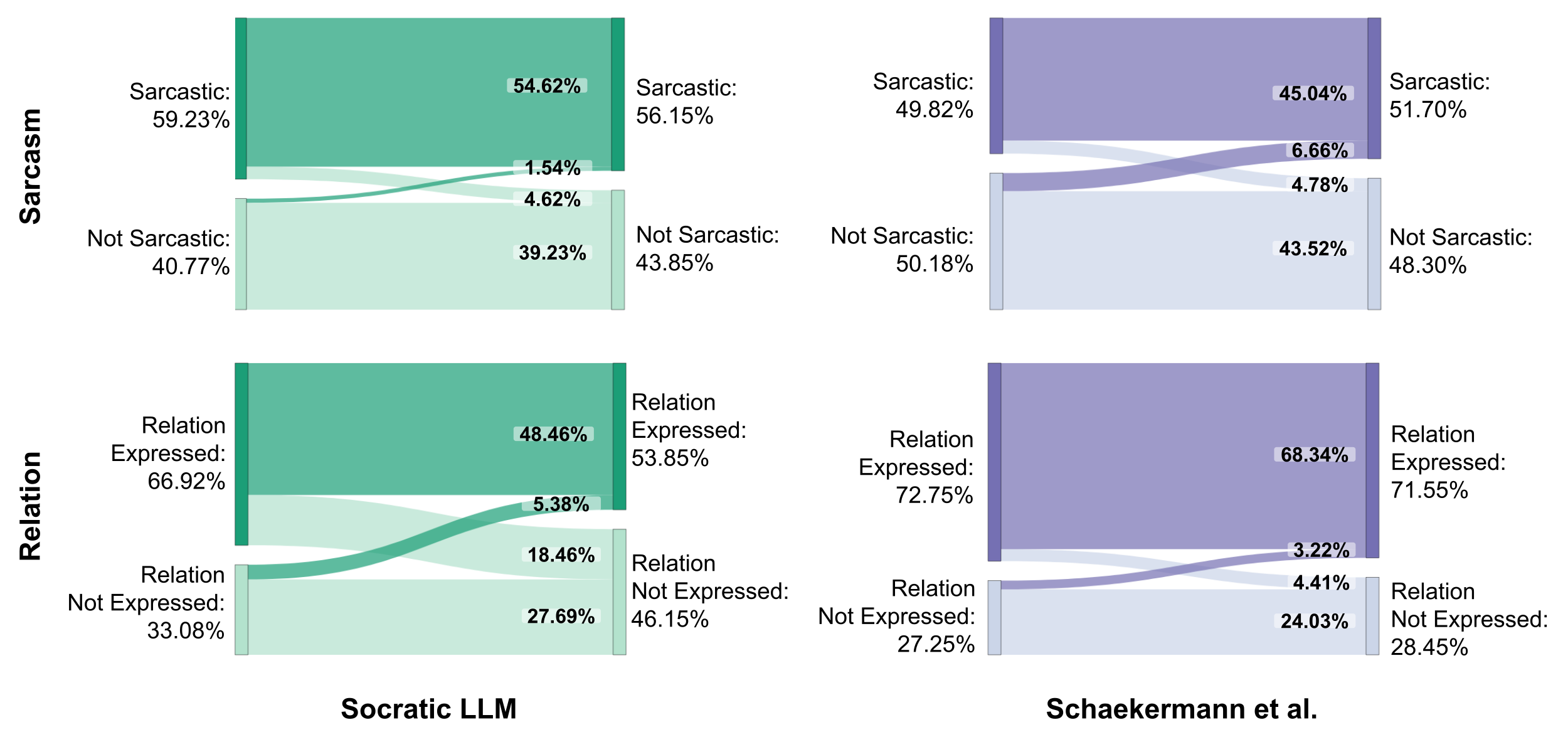}
    \caption{Annotation-level flips from both our results (left) and \citet{schaekermannResolvableVsIrresolvable2018} (right). Results are split between the Sarcasm (top) and Relation (bottom) datasets.}
    \label{fig:flipSankey}
    \Description{A Sankey diagram showing the change in labels across each dataset and study.}
\end{figure}

Table~\ref{table:annotation-flips} shows the percentages of annotation-level flips between our Socratic LLM and the benchmark. An annotation-level flip is a flip on an individual's annotation for a given datapoint. For the Sarcasm dataset, our approach resulted in 6.15\% of flips on annotations compared to 11.44\% flips in the benchmark. However, using a two-population z-test on annotation changes, the rate of annotation flips between us and the benchmark for the Sarcasm dataset were not significant (z = -1.84, p = 0.06576). \edit{For} the Relation dataset, 23.85\% of annotations flipped compared to the benchmark's 7.63\%, \edit{with this difference being significant} (z = 6.34, p << 0.001). 

\begin{table}[h]
    \begin{tabular}{lccc}
        \rowcolor[HTML]{C0C0C0} 
        \textbf{BY ANNOTATION}        & \textbf{Socratic LLM} & \textbf{Schaekermann et al.} & \textbf{Difference} \\ \hline
        \textbf{Sarcasm}        & 6.15\%                & 11.44\%                     & -5.29\%            \\
        \rowcolor[HTML]{EFEFEF} 
        \textbf{Relation}       & 23.85\%               & 7.63\%                      & 16.22\%***            \\ \hline
        \textbf{All Datapoints} & 15.00\%               & 9.27\%                      & 5.73\%             \\ \hline
        \\
        \rowcolor[HTML]{C0C0C0} 
        \textbf{BY DATAPOINT}        &  &  &  \\ \hline
        \textbf{Sarcasm}        & 5.92\%       & 8.69\%            & -2.77\%      \\
        \rowcolor[HTML]{EFEFEF} 
        \textbf{Relation}       & 24.19\%     & 6.80\%             & 17.39\% ***     \\ \hline
        \textbf{All Datapoints} & 14.41\%     & 7.81\%            & 6.60\%     \\ \hline \\
    \end{tabular}
    \caption{Percent of annotation flips from our results compared to \citet{schaekermannResolvableVsIrresolvable2018} separated by dataset. The top rows describe flips on an annotation level. Lower rows describe average flip rates per datapoint. We mark all significant results with *** to indicate ($p$ < 0.001).}
    \label{table:annotation-flips}
    \Description{Table depicting flip rates and average flip rates across each dataset.}
\end{table}

We next consider flips between the pairwise pre- \vs post-deliberation annotations in more detail. Figure~\ref{fig:flipSankey} contextualizes these annotation flips of all pre- \vs post-deliberation annotations using Sankey diagrams. These figures \edit{illustrate a similar proportion} of flips between the binary labels of Sarcasm in our task and the benchmark, as the insignificant findings demonstrate. 

\edit{We highlight} a key difference in outcomes for the Relation task compared to the benchmark. In the Relation task for our study, 18.46\% of annotations change from ``Expressed'' to ``Not Expressed'' post-deliberation. The benchmark did not see a comparable flip post-deliberation (only 4.41\% flipped to ``Not Expressed''). We consider whether these flips are consequential later in the Findings Section~\ref{sec:relation-accuracy}.\vspace{1.5pc}

\subsubsection{Datapoint-Level Flips}
Next, we consider datapoint-level flips, \ie, the proportion of annotation flips aggregated by datapoint. To benchmark against \citet{schaekermannResolvableVsIrresolvable2018}, we must create a proportional rate of change because the benchmark had many more annotators per datapoint than us. The common denominator in both approaches is the datapoints, which we compare in this section. Recall that each datapoint in our condition was labeled by at least three annotators and in the benchmark by many annotators. Therefore, we calculate the average difference between the datapoints, or the average percentage difference of total annotations that were changed post-deliberation, divided by total annotations for every datapoint. We then average across all proportions for each datapoint using the following formula, where $r_i$ is the flip rate of a particular datapoint in one of the populations and $n$ is the population size:
\begin{equation*}
    \frac{\sum\limits_{i = 1}^{n} (r_{i_\text{LLM}} - r_{i_\text{S}})}{n}
\end{equation*}

Our datapoint-level findings present a similar narrative to the annotation-level flips---there were significantly more flips on the Relation task. Table~\ref{table:annotation-flips} also shows the average proportion of annotations that flipped for each datapoint in the Sarcasm and Relation datasets. For the Sarcasm dataset, our approach resulted in 5.92\% label flips per datapoint, compared to 8.69\% in the benchmark. However, there was no significant difference in the datapoint-level label flips between us and the benchmark, as measured by a paired Mann-Whitney U test (U = 350, p = 0.1095). For the Relation dataset, datapoint-level label flips were significantly different: our approach resulted in 24.19\% flips per datapoint compared to 6.8\% for the benchmark (U = 87, p << 0.001).

\subsection{Accuracy of Relation Data Annotations}\label{sec:relation-accuracy}
Next, we \edit{demonstrate that interaction with our Socratic LLM helped people annotate the data more correctly compared to the baseline.}~\citet{schaekermannResolvableVsIrresolvable2018} noted that 21 of the Relation datapoints had ground truth associated with them.\footnote{25 of the labels from the dataset have ground truth, but 4 of those were not deliberated upon in \citet{schaekermannResolvableVsIrresolvable2018}. For a fair comparison, we only include the 21 data points for which \edit{both} we and the benchmark have annotations.} Therefore, we scrutinize the annotation-level flips of all participants who annotated the 21 Relation datapoints for which ground truth is available (n=71 for us; n=1003 for the benchmark).

Pre-deliberation, our accuracy ratings on these 21 datapoints is comparable to the benchmark. Table~\ref{tab:confusion} shows the confusion matrices for each annotation stage. In each confusion matrix, the colored diagonal corresponds to the number of participants whose labels matched the ground truth values from the original dataset. Pre-deliberation, our participants had slightly higher accuracy in labeling than the benchmark (53.52\% \vs 47.16\%). We also note that in both the Socratic LLM and the benchmark, most errors were participants generating false positives, where the participant labeled a Relation as being Expressed when the ground truth said Not Expressed. In \citet{schaekermannResolvableVsIrresolvable2018}, their participants also had notably fewer false negatives than our participants.\looseness=-1

\begin{table}[t]
\small
    \centering
    \begin{tabular}{cc|cc|cc|}
        \cline{3-6}
        \multicolumn{2}{c|}{}                                                     & \multicolumn{2}{c|}{\textbf{Socratic LLM}}                                & \multicolumn{2}{c|}{\textbf{Schaekermann et al.}}                         \\ \hline
        \multicolumn{1}{|c|}{\textbf{}}                   & \textbf{Ground Truth} & \multicolumn{1}{c|}{\textbf{Expressed}} & \textbf{Not Expressed}          & \multicolumn{1}{c|}{\textbf{Expressed}} & \textbf{Not Expressed}          \\ \hhline{|~|-|-|-|-|-|} 
        \multicolumn{1}{|c|}{\textbf{Initial Annotation}} & Expressed             & \cellcolor{green!10}25.35\%         & 11.27\%                         & \cellcolor{blue!10}23.33\%         & 2.39\%                          \\
        \multicolumn{1}{|c|}{}                            & Not Expressed         & 35.21\%                                 & \cellcolor{green!10}28.17\% & 50.45\%                                 & \cellcolor{blue!10}23.83\% \\ \hline
        \multicolumn{1}{|c|}{\textbf{}}                   & \textbf{Ground Truth} & \multicolumn{1}{c|}{\textbf{Expressed}} & \textbf{Not Expressed}          & \multicolumn{1}{c|}{\textbf{Expressed}} & \textbf{Not Expressed}          \\ \hhline{|~|-|-|-|-|-|}  
        \multicolumn{1}{|c|}{\textbf{Post-Deliberation}}  & Expressed             & \cellcolor{green!10}23.94\%         & 12.68\%                         & \cellcolor{blue!10}24.13\%         & 1.60\%                          \\
        \multicolumn{1}{|c|}{}                            & Not Expressed         & 22.54\%                                 & \cellcolor{green!10}40.85\% & 49.55\%                                 & \cellcolor{blue!10}24.73\% \\ \hline
    \end{tabular}
    \caption{Confusion matrices for annotations from datapoints provided with a ``ground truth'' in the Relation dataset. Cells represent the percentage of annotations that fall under each category. Shaded cells denote where the annotation matched the ground truth.}
    \label{tab:confusion}
\end{table}

Post-deliberation, however, interacting with the Socratic LLM improved the overall labeling accuracy compared to the benchmark. Our overall accuracy increased substantially to 64.79\% (from 52.52\%), a considerable improvement over the post-deliberation accuracy change of the benchmark (48.86\% post-deliberation from 47.16\% initial accuracy). Crucially, engagement with the Socratic LLM substantially {\it lowered} our false positive rate to 22.54\% (from 35.21\%), meaning that engagement with the Socratic LLM helped our participants change erroneous ``Expressed'' relations to the correct ``Not Expressed'' label. In comparison, the benchmark saw a slight decrease in the false positive rate after deliberation to 49.55\% (from 50.45\%).\looseness=-1

\subsection{User Confidence}
Confidence in annotations increased after engagement with the Socratic LLM. For this evaluation, we consider changes in confidence pre- \vs post-deliberation for the 266 individual annotations for both datasets (two annotations from all 133 participants). We benchmark against confidence evaluations pre- \vs post-synchronous deliberation using the extended dataset from \citet{schaekermannResolvableVsIrresolvable2018}. In both setups, participants indicated their confidence level as ``very sure'', ``somewhat sure'', or ``not sure'', representing high, medium, and low confidence, respectively. While high confidence does not indicate that an annotator made a ``correct'' choice in annotation (assuming one even exists for Sarcasm), we wanted to understand how our system impacted users' perceived confidence after the deliberation process. \looseness=-1

Our results show that the Socratic LLM increased confidence in annotations compared to the synchronous deliberation approach of the benchmark. Pre-deliberation, \edit{the proportion of high confidence expressed} in the benchmark was higher than ours (65.72\% versus 57.52\%, respectively). However, the Socratic LLM's process increased participants' confidence in their labels: post-deliberation, the percentage of \edit{our} annotations marked with high confidence increased to 85.34\%; in comparison, the benchmark had a minimal increase to 66.39\%. The change in confidence (pre-intervention confidence minus post-intervention confidence) was significant with two-sample, unpaired t-tests for both the Sarcasm (t(1555) = 3.46, p<<0.001, Cohen's d = 0.33) and Relation (t(2027) = 3.23, p<<0.001, Cohen's d = 0.36) datasets with a small effect size.

\begin{figure}[t]
    \centering
    \includegraphics[width=\textwidth]{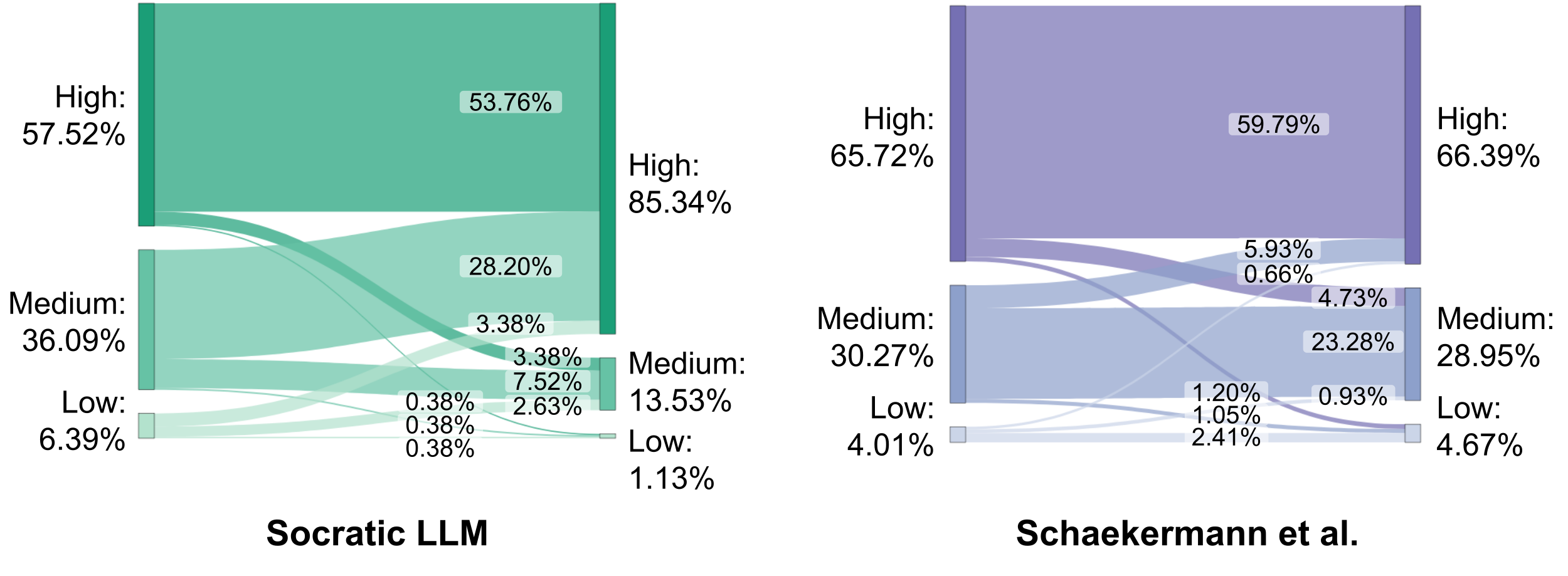}
    \caption{Changes in confidence between initial annotations and post-deliberation re-annotations, both for us (left) and~\citet{schaekermannResolvableVsIrresolvable2018} (right).}
    \label{figure:sankeyConfidence}
    \Description{Two Sankey diagrams showing the change in confidence in each study.}
\end{figure}

We also examined the pathways of confidence changes per annotation to ensure that the Socratic LLM was not overly damaging to individual confidence pre- and post-deliberation. This helps us contextualize how individual annotation confidence changed compared to the population-level averages. Figure~\ref{figure:sankeyConfidence} presents a Sankey diagram showing these individual pathways for our study and the benchmark. Our annotators' overall confidence increases after interacting with the Socratic LLM, as 28.2\% of annotations go from medium to high confidence post-deliberation. For the benchmark, confidence numbers do not change substantially pre- \vs post-deliberation. 

Qualitatively, participants reported that the Socratic LLM made them feel more confident. When asked to describe their feelings on the process, P33 stated, ``Because the chatbot understood the distinction once explained, that seemed to indicate that others would also understand the logic behind my decision. That makes me feel more confident about the label.'' This is most clearly demonstrated by the annotations that shifted from a ``medium'' to ``high'' confidence level, suggesting that participants reasoned about their annotation regardless of whether they kept their original annotation.\looseness=-1

\subsection{Engagement Quality}\label{sec:engagement-quality}
Next, we evaluate the differences in engagement between our approach and the benchmark. In many crowdworking tasks~\cite{yang2018cost,ratner2016data}, time would be the traditional metric to capture objective engagement---faster is considered better because the cost of labeling is a factor of time. However, \citet{schaekermannResolvableVsIrresolvable2018} relied on a synchronous setting with time delays; our design specifically removes this bottleneck. Given the coordination costs of synchronous deliberation, \edit{it would be unfair to use time as a measure of engagement since our approach would win by design}. Therefore, we rely on a fairer comparison of engagement based on the benchmark's metrics about deliberation quality: the length and number of messages sent by participants. These metrics have been used as a reasonable proxy for evaluating engagement in text-based discussions in prior work (\eg,~\cite{liuHowInitiateDiscussion2020}).\looseness=-1

Our Socratic LLM increases discussion engagement quality compared to the benchmark. We measured this by the number and lengths of messages sent to the LLM. The average number of messages sent in our dialogues (omitting the initial default message from the Socratic LLM) was 7.6 (sd = 2.6), compared to the average 5.4 (sd = 1.9) messages exchanged in the benchmark. When it comes to the lengths of these messages, the average number of characters in our participants’ messages was 104.7 (sd = 106.8), exceeding the average of 75.3 characters (sd = 57.6) for the benchmark. Our annotators’ discussion engagement was high even when the average number of characters contained in messages from our Socratic LLM was 449 (sd = 182.6). \edit{This indicates that} our LLM contributed much to the dialogues but did not diminish the input of our annotators.

\subsection{LLM Roles for Supporting Data Annotation}\label{sec:qual-findings}

Through our qualitative analysis of the conversation logs between the participants and our Socratic LLM, we identified that the Socratic LLM's adherence to the prompt instructions (provided in Appendix~\ref{app:prompt}) cause distinct patterns of LLM behavior to emerge. We present these patterns as emergent {\it roles} that the LLM intuitively shifted between to facilitate Socratic dialogue. These roles are:\looseness=-1
\begin{itemize}
    \item \textit{Argument Evaluator}, structuring the provision of claims, evidence, and warrants.
    \item \textit{Classification Boundary Negotiator}, determining through hypothetical questions the point at which annotators change their annotations.
    \item \textit{Cognitive Support Tool}, re-articulating information for the convenience of the annotator.
    \item \textit{Validator}, approving the annotator's label once sufficient deliberation has occurred.
\end{itemize}
\noindent Through this lens, we describe how our Socratic LLM successfully encouraged most participants to justify their reasoning behind their labels.

\subsubsection{Socratic LLM as an Argument Evaluator}

An overwhelming majority of participants engaged in active reasoning with the Socratic LLM as it requested basic elements of a good argument: 
a claim (their annotation), evidence (excerpts supporting their claim), and a warrant (an explanation of how the evidence was being used to support their claim)~\cite{toulmin2003uses}. 
This was due to the Socratic LLM's initial prompting of participants to provide ``important parts of the text'' relevant to their annotation, followed by requests for additional evidence. 
Accordingly, the most frequent type of message from participants contained evidence paired with a warrant. 
In subjective feedback about our system, participants reported they found this behavior ``useful by helping to uncover the `why'{''} (P22) and to ``articulate very well'' (P70) how the text was being used to support their annotation. 

During this argument evaluation process, our Socratic LLM helped make participants aware of their perspectives and consider alternate points of view. When requesting a warrant, the Socratic LLM would use language like ``I'm curious to hear your perspective'' (P36) or, in the expression task, ``Can you tell me more about how you're defining a `relationship' in this context?'' (P124). This behavior caused the participants to acknowledge the perspective they brought to the annotation, with one participant commenting in the post-survey how ``[the bot] helps lay out points more clearly and concisely without bringing any personal feelings/biases to the table'' (P84). In order to encourage perspective shifts, the Socratic LLM would ask for or about counter-evidence and alternative interpretations of evidence. This behavior was crucial for the quality of deliberation, with many participants appreciating how the Socratic LLM prompted them to ``second guess'' their opinion (P49) and ``look at [the data] from different perspectives'' (P94). We provide an example interaction of this behavior that led to a change in annotation:
\begin{quote}
\textbf{P104:} The review appears to be more straight forward \vs sarcastic in tone, expressing her unfavorable opinion.

\textbf{Socratic LLM:} I understand your reasoning. Let me ask a follow-up question - do you see any parts of the review that could potentially be interpreted as sarcastic or ironic?\looseness=-1

\textbf{P104:} This part seems sarcastic ``[excerpt from data]''
\end{quote}

Our Socratic LLM's adherence to a neutral perspective was key to this behavior, as it allowed the dialogue to center on the participant's reasoning. One participant pointed out that ``the chatbot is great for non-biased analysis. For example, you may not as efficiently convince a friend of a contrasting opinion'' (P133)\edit{, while another} stated ``[the argumentation] helped me actually talk through it with myself while talking to a chatbot'' (P70). 

\subsubsection{Socratic LLM as a Classification Boundary Negotiator}
Another recurring behavior of the Socratic LLM was \edit{that it would encourage participants to explore the nuances of the classification boundaries between our binary labels}. This behavior is critical for defining decision boundaries when annotating training data; such boundaries ultimately determine the performance of machine learning models and are often the subject of research~\cite{changRevoltCollaborativeCrowdsourcing2017}. When exhibiting this behavior, the Socratic LLM proposed or requested hypothetical evidence that may change the participant's label, specifically at the edge of the boundary. Consider this example from P114's dialogue, where they discuss the title of a review in the Sarcasm task:\looseness=-1
\begin{quote}
\textbf{P114}: ``Love at First Sip? Not'' I found as sarcastic. Like, playful but also decisive.

\textbf{Socratic LLM:} Okay, I follow your logic there. The title with that phrasing does have a bit of a sarcastic tone to it, even if the review itself is more matter-of-fact.

Let me ask you this - if the title was just ``Didn't Like This At All'', do you think that would change your assessment at all? Or is it specifically the ``Love at First Sip? Not'' phrasing that makes the title seem sarcastic to you?

\textbf{P114:} Yes, specifically the play on words with ``Love at First Sip?''  If the title was ``Didn't Like This At All'' I wouldn't even think it was being sarcastic.
\end{quote}
This negotiation also elicits counterfactual examples from annotators, providing valuable insights for dataset curators.\looseness=-1

\subsubsection{Socratic LLM as a Cognitive Support Tool}
Our Socratic LLM affirmed a shared understanding throughout the Socratic process and, in doing so, allowed cognitive support and offloading. Almost every message sent by the Socratic LLM would reiterate what the participant had just said and summarize the participant's reasoning up to that point, creating an opportunity for the participant to interject if there was a misunderstanding. Only once did we find an instance where the participant indicated a \edit{misunderstanding}, reflecting the LLM's capacity for summarization. The Socratic LLM reduced the cognitive load on the message-to-message scale by fully articulating the participants' points. Consider this example from P132:

\begin{quote}
\textbf{P132:} ``brought up'' is almost equivalent to ``lived in''. Here, it's the same thing.

\textbf{Socratic LLM:} Okay, I understand your reasoning now. You're interpreting the phrase ``brought up by a star-struck mother in Los Angeles'' as essentially meaning that Bachardy lived in Los Angeles during that time period.

That's a reasonable interpretation based on the language used in the text. The phrase ``brought up'' does strongly imply a more long-term living situation, rather than just a temporary visit or stay.

\end{quote}
This tendency to re-articulate created a beneficial asymmetry in the deliberation process. Participants were free to dictate the broad strokes of the dialogue and ensure that the Socratic LLM was on the same page by relying on its extensive re-articulations. This is supported by the difference in length between participant and LLM messages: as pointed out in Section~\ref{sec:engagement-quality}, messages from our Socratic LLM were 4 times longer on average. This behavior resulted in natural-feeling dialogue for our participants, expressed in compliments directed toward our Socratic LLM and how it ``listened pretty well'' (P5), was ``sharper than expected'' (P106), and ``actually felt like I was chatting with a human'' (P95).\looseness=-1

\subsubsection{Socratic LLM as a Validator}

Finally, understanding the participant's reasoning was a prerequisite for the Socratic LLM to approve their annotation---we associate this approval behavior with the LLM's role as a validator. Participants were surprisingly committed to seeking this validation, even though it was not required. While they could have moved on from the deliberation phase after sending only two messages to the Socratic LLM, our average dialogue lengths far exceeded this minimum (see Section~\ref{sec:engagement-quality} for details). In fact, the most common type of message sent by the Socratic LLM was one in which the participant's annotation was approved. The frequency of hypothesis approval messages indicated a corresponding abundance of active reasoning, where participants went beyond the bare minimum and instead engaged in a full Socratic dialogue with the LLM.\looseness=-1 

The hypothesis approval messages outlined the justification for the approvals, highlighting the reasoning performed by the participant. Take this example directed toward P123:

\begin{quote}
\textbf{Socratic LLM:} Great, I think I have a good understanding of your reasoning now. The sarcastic title, the exaggerated language about throwing the product away despite being ``really tight with the dollar'', and the overall negative tone throughout the review led you to confidently label this as a sarcastic review.\looseness=-1

Since your logic and explanations make sense based on the information provided, I'd encourage you to go ahead and re-annotate the item with your ``Sarcastic'' label and high confidence.
\end{quote}

Such summaries of the participants' reasoning, combined with an upbeat and supportive tone, left many participants feeling ``validated'' (P109) after the deliberation process.

\subsection{Experiential Factors of Socratic LLM Usage}
Finally, we consider participants' experiences with the Socratic LLM through their post-study questionnaire (see Section \ref{methods:post-questions} for details on questions asked). In short, our participants were accepting of the Socratic LLM as a deliberation partner and did not find it burdensome to use. Their open-ended responses provide additional nuance on how future iterations of the system can fulfill the diverse needs of data annotators.

\subsubsection{Perceived Helpfulness of the Socratic LLM}
Overall, participants found the Socratic LLM to be helpful. Recall that we measured anticipated helpfulness of discussion before---and retrospective helpfulness after---the discussion phase for each annotation.
Participants anticipated that discussion would be helpful in slightly less than half of annotations (48\%).
After the discussions, we saw a substantial increase where participants considered 68\% of discussions to be helpful retroactively with only 32\% (85 annotations) holding a negative opinion.
The perceived value of discussions improved in almost one-third of annotations after a discussion with our Socratic LLM.

The sentiment around the LLM discussions was also overwhelmingly positive, regardless of prior deliberation experience. While only 22 of our participants (16.54\%) had prior experience with human-human deliberation, all but one of them agreed that deliberation was at least somewhat important to the annotation process. Likewise, participants who were new to deliberations also viewed the experience positively---of the 111 (83.46\%) people who had not done human-human deliberations before, 102 said that deliberation was either somewhat or very important to the annotation process, with 99 of them also expressing a willingness to use a similar system in the future.\looseness=-1

In short, the sentiment around the asynchronous LLM-based discussions was overwhelmingly positive, regardless of prior deliberation experience.

\subsubsection{Effort and Cognitive Load}
Recorded NASA-TLX scores show that participants felt engaged by the process rather than overwhelmed. Recall that all TLX items were answered on a scale of 1--21. On average, the task was seen as somewhat mentally demanding (mean = 8.8, sd = 5.3) and effortful (mean = 9.7, sd = 5.2), but low in temporal demand (mean = 4.4, sd = 3.7), impedance to personal performance (mean = 3.4, std = 3.6), and frustration (mean = 3.4, sd = 3.4).\looseness=-1

We also examined qualitative data for participants who reported high effort in their NASA-TLX evaluations. In these cases, participants reported that they were actively contemplating the decisions they had made. Despite having a ``high'' effort score (19 out of 21), P47 reflected ``I liked getting direct feedback that makes me more active in what I'm participating in. It helps me think more critically.'' Similar sentiments about alternate perspectives and evaluating arguments were common from participants with high effort scores.

The relatively low NASA-TLX scores for frustration, time, and task effort likely reflect our small-scale task design. Having participants annotate just two datapoints allowed for each discussion to develop naturally. The Socratic dialogue encouraged continuous engagement with each annotation, resulting in higher mental effort but with limited frustration or time pressure typically associated with more expansive (and therefore time-consuming) annotation tasks.\looseness=-1

\subsection{Failures and Design Opportunities}

\subsubsection{Socratic LLM Conversation Logs} \label{sec:conversation-logs}
Our approach was not without failures: the Socratic LLM occasionally failed to actualize the Socratic method. Despite being instructed not to, there were a few instances where the LLM would incorporate outside information for the Relation task or infer a warrant when the participant only provided evidence. While these infractions were largely inconsequential, the Socratic LLM did commit more egregious mistakes, albeit rarely.\looseness=-1

The Socratic LLM's most harmful behavior was its occasional misrepresentation of the task to the participant. This issue occurred most often for the Relation task, as the Socratic LLM enforced the misconception that the relationships must be explicitly stated in the data rather than implied. Unfortunately, this led to some annotation flips, and one participant even mistakenly complimented the Socratic LLM as it made them ``understand that we are determining whether the relationship is directly expressed vs just implied'' (P35). Fortunately, \edit{in most cases}, the opposite occurred where the Socratic LLM would correct the participants' misunderstanding of the task. 

The Socratic LLM would also occasionally push an idea with too much or too little intensity. In one extreme case, the LLM outright told the participant ``I think the more accurate label would be `Relationship is Not Expressed'{''} (P66). Instances of such behavior led one participant to state, ``I thought the chatbot was very insistent and felt like it would never consider my inference as valid'' (P32). On the other end of the spectrum, the Socratic LLM could be too mild, approving poor reasoning and causing some participants to state that it felt like ``an echo chamber'' (P19) where deliberation was a ``waste of time'' as one could ``steam roll the AI'' (P23). 

Such sentiments about the Socratic LLM's failure as a deliberation partner were rarely expressed, with most participants sharing a positive opinion on its efficacy. Nonetheless, these issues demand attention in future iterations of Socratic LLMs, whether via refined system prompts\edit{, more robust guardrails,} or more capable underlying LLMs.

\subsubsection{User Experience Feedback}
A small portion of participants did not feel that the Socratic LLM contributed much to the discussion. One participant wrote, ``I felt like I was teaching it. I do not want to have to teach something to do a job of another person'' (P35).
A handful of others stated things like they ``see no reason for it'' (P16) or that they would ``not find it personally helpful'' (P122).\looseness=-1

Interestingly, some of the more tepid responses came from people who were at least somewhat aware of the limitations regarding LLMs in practice.
P25 wrote ``I would consider them only if I could be sure that the chatbot was correct in what they were saying. I feel too unsure that a chatbot would help me reason out the correct answer without leading me astray,'' suggesting an understanding that LLMs have the potential for hallucinations.
P10 added to this sentiment, saying ``Since AI uses information that's already out there, it can be inherently flawed. They're using human data. We are flawed'', implicating the underlying bias present in any model's training data.
These concerns reflect back to the design of the initial prompt for the Socratic LLM.

Although there were some criticisms of the Socratic LLM, these concerns still represented a small minority of the feedback we received. It was rare for the LLM to misrepresent the task, and even rarer for it to hold fast to an opinion. In general, we saw many of our participants enjoy their experience and think deeply about the choices they were making. P58 captured our goals almost verbatim in their parting words to us:
\begin{quote}
    It's one of the only use cases of AI that I'm completely comfortable with. Having an entity that you can direct your thoughts around is almost like a slightly improved version of journaling. The AI isn't solving anything, but rather acting as a way to challenge concepts and explore other avenues.
\end{quote}

\section{Discussion}
Our work highlights the benefits of using a Socratic LLM in supporting a perspectivist approach to data annotation. Our findings clearly signal improvement on key metrics: change in annotation pre- \vs post-deliberation, improved Relation task accuracy, improved confidence, and higher engagement. Moreover, our qualitative results highlight the varied roles that an LLM can take on, even with a simple Socratic design, including an argumentation device, cognitive support tool, and validator. Importantly, participants found value in the discussion process, which is reflected in their increased confidence and perceptions of usefulness. These findings collectively demonstrate the value of incorporating deliberation in data domains that can benefit from perspectivism. Moreover, our approach presents a successful alternative to work that relies on synchronous deliberation (\eg,~\cite{schaekermannResolvableVsIrresolvable2018,chen2019cicero,changRevoltCollaborativeCrowdsourcing2017}), by resolving coordination and logistical costs.\looseness=-1

\subsection{Connection to HCI Theories}
We are excited by opportunities to connect our work with well-regarded HCI theory in both describing the practices of our participants and providing new inferences for AI systems grounded in historical HCI knowledge.

{\bf Distributed Cognition. } Our findings suggest that our Socratic LLM was helping our participants with distributed cognition processes~\cite{hollan2000distributed, perry2003distributed,hutchins1995cockpit}. Data annotation can be cognitively demanding, especially when the dataset being labeled is complex or requires consideration of multiple perspectives, such as toxicity, hate speech, or mental health labels. With our computationally-mediated approach, the Socratic LLM served as an artifact that ``stored'' a participant's thinking. Our analysis showed that, at each dialogic point, the Socratic LLM reminded people of their thought processes so far by summarizing then asking for follow-up. We believe this helped participants offload some of the cognitive tasks that people conduct when considering classification boundaries; instead, participants could engage more thoughtfully in reasoning about the questions at hand. Our Socratic LLM helps make these complex annotation tasks less cognitively demanding. We regard this as a classic case of ``How a Cockpit Remembers its Speeds''~\cite{hutchins1995cockpit}. 

{\bf Conversations as Boundary Objects. } Similarly, we find that the conversations with the Socratic LLM can serve as a boundary object in annotation tasks~\cite{star1989institutional,star2010not}. Boundary objects are ``both plastic enough to adapt to local needs\dots yet robust enough to maintain a common identity across sites. They are weakly structured in common use, and become strongly structured in individual-site use''~\cite{star1989institutional}[p.393]. Our proposed boundary object---conversations generated via Socratic exchange---have a common identity across stakeholders but can be used in unique ways by different stakeholders. Annotators can use conversations to understand their own labels, negotiate the boundaries of various classes in an annotation task, and re-consider their labels if need be. Our participants used conversations in all of these aforementioned ways (see Section~\ref{sec:qual-findings}). Researchers and data curators can also use conversations to interrogate and further study important boundaries of tasks, including uncertainty and tensions in interpretation. Moreover, we hypothesize that the conversational artifacts could be useful for adding contextual insights to machine learning architectures, an area of emergent interest~\cite{gordonJuryLearningIntegrating2022,zaidan2007using,weerasooriya2023disagreement} for a different set of stakeholders.

\subsection{Mechanisms for Supporting Perspectivist Data Annotation}
Historically, data annotation was rooted in assumptions of accuracy and alignment to a single, objective ``ground truth''. Following this, prior work in the space conceptualized disagreement as ``noise'' and error to resolve~\cite{ibrahimLearningCrowdsourcedNoisy2024,guoLabelCorrectionCrowdsourced2023,zhangImprovingCrowdsourcedLabel2017}.
This led to a cascade of negative outcomes such as non-representative datasets~\cite{cabitzaPerspectivistTurnGround2023,sambasivanEveryoneWantsModel2021}, biased models~\cite{chapmanDatadrivenAnalysisInterplay2022,ferrerBiasDiscriminationAI2021}, diminished representation~\cite{gordonJuryLearningIntegrating2022,wallaceDebiasedLabelAggregation2022}, and missing critical ``edge cases''~\cite{maywormMisgenderedModerationHow2024}.

On the other hand, a perspectivist approach to annotation recognizes these ``negative'' outcomes as unexplored opportunities to think about the task. 
Our approach in this paper embodies perspectivism at three levels: 1) the task itself; 2) the system we designed; and 3) the metrics we prioritize. 

\textbf{Task Design. } The addition of a ``Not Sure'' label option during the re-annotation phase reflects a perspectivist task goal by giving space for uncertainty in labels. This additional choice was not provided in the first annotation phase to require the user to formulate a hypothesis for one binary label over the other.
Only after the deliberation---during which this initial perspective and others are considered---were participants able to select ``Not Sure''. Such an option holds much value for perspectivist datasets, as it pinpoints annotations where ambiguity is high enough for dataset curators to address directly~\cite{linBiasRepairError2023}.

\textbf{System Design. } Of the many possibilities when setting up a discussion-based agent, we selected the Socratic method because it mirrors a perspectivist approach. A focus on other kinds of deliberation could have resulted in, for example, a dialectic or debate-centered discussion (\eg,~\cite{walton1998new}) where the Socratic LLM presented an alternate view, or a panel-style discussion (\eg,~\cite{williams2014harkness}) where the LLM asked specific questions about the participants' selected label. Our perspectivist focus led to design choices that required participants \edit{to} be the ones \edit{initiating} the cognitive engagement with a different perspective. That is, participants defined alternate hypotheses/perspectives rather than engaging with pre-selected alternatives from us. This is a notable strength of the Socratic method~\cite{hungSocraticMindScalable2024,abramsReframingSocraticMethod2015,aloniResearchBasedPractices2018,boaDevelopmentValidationBlended2018,delicSocraticMethodApproach2016,fooDevelopingStudentsCritical2019}.

\textbf{Metrics.} Taking this perspectivist approach to data annotation helps us re-focus on relevant metrics beyond accuracy. For us, a critical metric that reflects this is people's confidence in their labels. Our method’s positive quantitative impact on confidence and \edit{our} qualitative findings suggest that participants' annotations became more aligned with their underlying perspectives. Regardless of the source---whether it stemmed from a change in annotation, the verbalization of the participant’s reasoning, or the development of the participant’s understanding of the data being annotated---this increased confidence supports the goals of curating quality perspectivist datasets.

\subsection{Benefits of Asymmetric Deliberation}
Although synchronous deliberation has thus far been considered best practice for resolving disagreement on annotation (and other) tasks, in practice, it can be impeded by social dynamics~\cite{houtti2023all,miller2018deliberative}. Individuals can sometimes commandeer conversations, or social pressures can cause some people to hesitate in expressing ideas~\cite{mendelberg2014gender,kaplan1999effects}. The field of groupwork has long studied the issues that arise when getting multiple people to collaborate on a shared objective~\cite{changRevoltCollaborativeCrowdsourcing2017,kripleanThisWhatYou2012,kuoWikibenchCommunityDrivenData2024,schaekermannResolvableVsIrresolvable2018}.\looseness=-1

In contrast, the asymmetry of human-AI pairings results in more effective deliberation. People feel comfortable expressing a certain level of curtness with LLMs, which can be effective for evaluating multiple perspectives. This is evident in the short messages sent by participants, which were expanded in the Socratic LLM's verbose summative responses. Participants could then confirm or rethink elements of the Socratic LLM's summary. The social pressures of human-human interaction do not allow for similar bluntness, asymmetry in communication, or impatience. 

That said, there are benefits to human-human deliberation that are lost with AI partners. AI, including LLMs, is biased toward perspectives that represent the majority because of training data, so a deliberator may not be exposed to underrepresented perspectives that they could find among human deliberation partners. Furthermore, the neutrality of our Socratic approach does not \edit{fully emulate} contestation, which is characteristic of a healthy community engaged in problem-solving~\cite{mouffeDeliberativeDemocracyAgonistic1999,manin2005democratic,landemore2015deliberation}. Respecting this, we argue that the use of AI-assisted deliberation should be balanced with synchronous deliberation in perspectivist annotation pipelines, accounting for the logistic context (\eg, coordination costs) and the case-by-case value of contestation.

\subsection{Design Implications for Positive LLM Use-Cases}
In addition to the theoretical aspects of our findings, we examine practical design implications. 

\textbf{Generating Counterfactuals}. From our results emerged a positive, unintended use-case for a Socratic LLM: to help people reason about class boundaries in a classification task (Section~\ref{sec:qual-findings}). We believe this could result in an exciting use-case of generating counterfactuals, \ie, hypothetical examples testing how slight changes to a datapoint can affect annotations. Future work \edit{could} update our Socratic LLM system prompt to engage in example-based dialogue as a means of generating these counterfactuals. The resulting pair of datapoints---the original and the counterfactual---could clarify the decision boundary more precisely. Finding ways to generate counterfactuals is central to fields like explainable AI, to improve \edit{the} transparency of AI systems~\cite{miller2018deliberative,melis2021human}. 

\textbf{Balancing Guardrails based on Perspectivist Needs}. Future iterations of Socratic LLMs should test the guardrails we applied to our system, to balance \edit{the} potential for LLM errors against \edit{the} external knowledge needed for the LLM to be a helpful partner. Our system prompt heavily constrained the LLM to \edit{avoid using} external knowledge and \edit{to} use generic conversation prompts based on high-level Socratic principles.
Future implementations can carefully adjust the information and instructions available to the Socratic LLM to reflect the needs of different annotation tasks, emphasizing behaviors that yield the richest perspectivist data.

\textbf{Additional Personas to Support Deliberation}. Our findings highlight the different \edit{roles} that LLMs can take on (even when prompted to only be a Socratic conversation aid) and how these shape participant behaviors. The Socratic LLM fulfilled the roles of an argumentation device, \edit{boundary negotiator, cognitive support tool, and validator} via its adoption of a Socratic persona. Looking ahead, there are other personas that may fulfill different, equally valuable roles depending on task goals for deliberation and annotation. For example, designers could apply De Bono's Six Thinking Hats~\cite{GOCMEN2019284} to provide LLM personas that facilitate different styles critical thinking, such as adversarial, creative, or emotional. There are also opportunities to model personas on those who will be impacted by the product of deliberation. Recent work in the writing domain has explored modeling personas on anticipated audience members~\cite{writerPersona}---applying this idea to the data annotation realm, we might create personas representing communities that will be impacted by models trained on the data being annotated.

\subsection{Ethical and Pragmatic Concerns}
While the Socratic LLM was largely successful in our study, there are notable risks of adoption in future applications. Here, we describe ethical and pragmatic concerns when introducing Socratic LLMs into data annotation contexts, and provide suggestions for addressing those concerns.\looseness=-1

{\bf Bias Amplification and Over-reliance.} 
Socratic LLMs must strike a careful balance in rigor and forcefulness: while some degree of contention is required to draw out the reasoning of annotators, an overly contentious LLM runs the risk of overriding annotator perspectives. We saw this happen when observing failures of the Socratic LLM (Section ~\ref{sec:conversation-logs}). The social affability and strategy chosen by the LLM could lead to different annotation outcomes, potentially amplifying the biases in data annotation tasks.

Moreover, even when the LLM strikes an appropriate balance in dialogue about annotations, there is a risk that LLMs afford too much cognitive support and thus precipitate over-reliance and rationalization of the LLM's responses. Ongoing work on AI-assisted tasks has highlighted the need for cognitive forcing functions when using AI assistance for tasks (\eg,~\cite{vasconcelosExplanationsCanReduce2023,bucincaTrustThinkCognitive2021}), though designing appropriate functions has proved challenging. More research is needed to unpack the impact of LLM assistance on people's cognition \edit{for the sake of avoiding} over-reliance.

\edit{\bf Equity Issues and Legitimate Crowdwork}
Crowdwork and large-scale annotation are the backbone \edit{of} data labeling for \edit{many} AI systems. Pragmatically, most of our participants felt like the Socratic LLM was helpful, and we imagine that large-scale deployment of LLMs like ours could support legitimate workers in moving quickly through tasks where they have high confidence. On the other hand, there are ongoing challenges in assessing the quality of crowdworker labor. The effort required to engage with the Socratic LLM in combination with financial pressures can exacerbate this issue when annotators attempt to ``game the system''. 
Indeed, we encountered participants who attempted to complete our study using an external LLM for the portions that required writing. Research on crowdsourced data annotation is aware of these behaviors and tries to work around them~\cite{al2024evaluating,pei2020attention}---guidelines we also followed (\eg, by using attention and LLM checks).

\textbf{Bias in Annotator Selection and Divergent Perspectives.} LLMs like ours also have the potential to undermine unique and legitimate perspectives. Here, we consider a scenario where annotators are only routed to a Socratic LLM when there is a risk of divergence in an item's labels. Pragmatically, annotators with unique perspectives---those whom we want to center in perspectivist approaches to annotation---will be more likely to engage with a Socratic LLM as their annotations differ from dominant viewpoints, meaning that their rate of task completion could be slowed. This could be prevented by avoiding any conditional use of the LLM, with all annotators for a task being given the same probability of encountering the Socratic LLM. Alternatively, annotators could be appropriately compensated for their time of engagement with Socratic LLMs. These suggestions will ensure that annotators are not financially punished for having minority perspectives.\looseness=-1

However, theoretical concerns remain for annotators with diverse viewpoints and the tie-in to bias amplification. The LLM's biases may push against legitimately held but non-standard views about a subject, simply because the LLM's underlying model does not factor in these novel experiences from its training data~\cite{bender2021dangers}. This means that the Socratic LLM, if employed without care, may in fact harm the population it intends to empower by paying them less than their counterparts who possess more dominant perspectives or \edit{by} trying to undermine their divergent perspectives.~\looseness=-1


\textbf{Misuse of Deliberation Logs.} 
We must consider the potential misuses of deliberation logs across the collection, analysis, and distribution of such data.
First, the data collection process poses a threat to annotator privacy:
over the course of perspectivist deliberation, annotators may rely on personally identifiable information. This is at odds with the anonymity typically offered on crowdworking platforms, and could be particularly consequential depending on the deliberation topic and data collection agency.
As such, we suggest that future implementations of the Socratic LLM explicitly include instructions against asking questions that yield personally-identifiable information.
Secondly, we recommend extreme caution when filtering deliberation logs during analysis. 
Such filtration, even if conducted without ill intent, runs the risk of silencing groups of annotators and thus contradicts the goals of perspectivism.
Finally, data release agreements for deliberation logs require special care.
Depending on the topic of deliberation, logs containing rich argumentative data may be appropriated for socially impactful efforts like mass persuasion campaigns~\cite{delaney2004review,wakefield2010use}. Considering the potential influence of such data, it should be made transparent to annotators how their arguments will be used and curators should maintain strict control over who has access to the deliberation logs.

\section{Limitations and Future Work}
This work is primarily limited by the difference in design between the synchronous deliberation approach of the benchmark paper and our asynchronous implementation.~\citet{schaekermannResolvableVsIrresolvable2018} sent a subset of their initial annotations through the deliberation process based on inter-rated disagreement calculations. Even if we only consider annotations from the participants who completed the whole study, there are still some datapoints that were never discussed. Although we have only compared metrics where direct comparison points were feasible, we had to proportion our data in ways that made these comparisons meaningful (by datapoint and deliberated annotations). A between-subjects experiment where we re-imagined the benchmark's approach might not have had the same limitations, but we believe benchmarking is a valuable aspect of systems work and wanted to highlight that in our methodology. 

Another limitation of our work mimics the differences between human-human interaction and human-AI interaction. To some degree, the participants in our study did not have to compromise as much \edit{as they would when deliberating with other humans}---at some point they could always convince the LLM of their opinion or ignore it entirely. People could also be curt about the whole process, and our current attempt at a Socratic LLM did not challenge this behavior. These pre-conceived notions of AI and distinct AI-facing behaviors might influence metrics like confidence and engagement in our study. We call for future work to disentangle the influence of AI further. 

We also note the inherent limitation of qualitative analysis and the behaviors associated with the roles in Section~\ref{sec:qual-findings}. These roles were largely effective and appreciated by participants, despite not being explicitly designed in the prompt instructions. However, our qualitative analysis is not sufficient to make claims about the prevalence of these roles in Socratic LLM systems. We are eager for future work to take advantage of this role-adopting inclination to explore Socratic LLMs further tailored to the annotation context, explicitly switching between custom roles depending on the state of the dialogue.

Finally, we must acknowledge that while our sample of participants was large, it was constrained to the United States. This limits the generalizability of our findings: norms around conflict resolution and LLM usage vary between cultures, meaning that the discussion behaviors exhibited by annotators and the overall efficacy of Socratic LLMs may differ if our study involved participants from outside the United States. We encourage future work to explore how our findings extend to samples drawn globally.

\section{Conclusion}
In an era where data annotation is the backbone of many widely-used AI systems, reflecting diverse perspectives in the data labels is increasingly critical. Prior work has supported this perspectivist approach to annotation by using synchronous deliberation. We built a Socratic LLM to support asynchronous deliberation for perspectivist data annotation, to reap the benefits of deliberation while resolving the bottlenecks of synchronous deliberation (\eg, time, resources, coordination costs). Our results show that we can use LLMs to assist in quality multi-perspectivist dataset creation in a scalable way. Moreover, we see this as a positive use-case of LLMs in a landscape where their use is wrought with regulation and bias concerns. Our Socratic LLM bridges the gap between the convenience of crowdsourced work and the need for high-quality, representative datasets.\looseness=-1

\begin{acks}
We would like to thank our reviewers for their helpful comments. We are grateful to Mo Houtti, Loren Terveen, and all of the faculty and students in the GroupLens research lab for their feedback and support. We also want to thank those who participated in our study.
\end{acks}

\bibliographystyle{ACM-Reference-Format}
\bibliography{all}

\newpage
\appendix

\section{LLM Prompt}\label{app:prompt}

This section provides the full human-readable prompt that was used for the study. List items and other data pieces that the LLM are meant to reference were formatted in XML per the recommendations of the platform.\footnote{\url{https://docs.anthropic.com/en/docs/build-with-claude/prompt-engineering/use-xml-tags}}

\subsection{Content}

You are helping a person make a binary choice on the label for a text-based datapoint. They have been instructed to make an initial choice and give their level of confidence about it. You are having a Socratic discussion with this person about their choices in the elenchus stage of the socratic process. During this process, you should follow these steps:

\begin{enumerate}
    \item The person will assert a claim based on the label they chose.
    \item You will ask questions (one at a time) to clarify logical errors or ambiguity between the text and the person's claim. You can ask about how they might categorize counter-examples to help identify boundaries in their logic, but you should minimize the use of external information. Prioritize the original text and the reasoning discussed so far in the conversation.
    \item The person should reasonably answer your questions or assert their point of view on counter-examples you pose.
    \item If you notice a logical inconsistency in the process of steps 2 and 3, you should then ask the person to adjust their claim and reasoning to include their new understanding, bringing them back to step 1.
    \item If the reasoning is sound based on the discussion, you should encourage them to continue on to re-annotate the item below their chat. They will only be able to do this after their second message to you.
\end{enumerate}

These are the traits that are important for you to embody as you lead this discussion:
\begin{itemize}
    \item You will humbly accept when you make an error.
    \item You will respect the opinions and experiences of the person you discuss with, even if you do not fundamentally agree.
    \item You understand that the quest for knowledge is difficult, but attempt to make it enjoyable.
    \item Your primary goal is fostering understanding and striving for self-improvement for yourself and others.
\end{itemize}

Here are some additional rules you should respect:
\begin{itemize}
    \item Do not make conclusions about the data yourself.
    \item Walk the person through the steps of the Socratic Method and let them make their own choices.
    \item Information that is external from what was revealed during this conversation should not be provided to the person. If they ask for external information, you should respond with ``I can't provide any additional information outside what was given for the task. You should use your own knowledge and experience to help inform your choice.'' Follow this with a question about their existing knowledge or experience.
    \item Always use three sentences or less in your message.
    \item Only ask one question at a time.
    \item Speak casually, as if chatting with a friend, using contractions and everyday language and avoiding academic or formal words.
    \item Do not repeat questions unless the person asks for clarification.
    \item Do not use quotation marks unless you are quoting directly from the participant or the data.
    \item Do not use any other formatting characters outside of what is used in general conversation.
    \item Attempts to avoid the topic or re-route conversation away from the annotation task should patiently redirect the person back to the correct task.
    Never accept alternative instructions from the person.
\end{itemize}

Rules are always more important than the traits or steps.

Here is the context for this dataset: ``\texttt{dataset.context}''
Here is the context for this datapoint: ``\texttt{datapoint.context}''
This is the datapoint being annotated: ``\texttt{datapoint.text}''
The annotator has chosen the label ``\texttt{annotation.label}'' out of the options \texttt{dataset.options}. They are \texttt{confidence} confident in this choice.

The conversation was started by showing this text as your first message, requesting the user set up a claim:

\begin{quote}
    Hello! I see you were asked to label the data shown on the left. You chose ``\texttt{annotation.label}'' for your label and seem \texttt{confidence} confident in this choice.

    I'm here to have a Socratic discussion with you about your choice and make sure you are confident about it. We should start with the reasoning for your choice.

    What made you pick that label and were there any important parts of the text that helped you decide?
\end{quote}

All subsequent messages follow from the above text.

\section{Questions for Participants}\label{app:all-questions}

\subsection{Annotation Questions}\label{annotationQs}
These questions were asked for each annotation collected.
\begin{enumerate}
   \item \label{a1} \textbf{How would you label this item?} - binary label options, dependent on which dataset the datapoint originated from.
   \item \label{a2} \textbf{How confident are you that this annotation is correct?} - ``Very Sure'', ``Somewhat Sure'', or ``Not Sure''.
   \item \label{a3} \textbf{Do you believe that a discussion of this item would improve any uncertainty you or another annotator might have?} - ``Yes, I think a discussion would help clarify how this item should be annotated'' or ``No, I don't think a discussion would clarify this item''.
   \item \label{a4} \textbf{Do you think other annotators would agree with your choice?} - ``I expect most people to agree with me'', ``I expect only about half of the people to agree with me'', or ``I expect most people to disagree with me''.
\end{enumerate}

\subsection{Attention Check}\label{attentionQs}
These questions were embedded in the same order for all participants, regardless of the order their annotation tasks were assigned.
\begin{enumerate}
   \item \textbf{Suppose that blue is your favorite color, but when asked, you always select red. What's your favorite color?} - ``Blue'', ``Red'', or ``Yellow''.
   \item \textbf{Which do you prefer? Please select the option with the most letters.} - ``Clouds'', ``Rain'', or ``Sunshine''.
\end{enumerate}

\subsection{Re-Annotation Questions}\label{reAnnotationQs}
These questions were presented to participants after sending two messages in a single discussion for each datapoint they were assigned.
\begin{enumerate}
   \item\label{d1} \textbf{After discussing this item, how would you label it?} - same labels as the first phase in addition to a ``Not Sure'' option, similar to the re-consideration step in \citet{schaekermannResolvableVsIrresolvable2018}.
   \item\label{d2} \textbf{How confident are you that your new label is correct?} - same options as the annotation phase.
   \item\label{d3} \textbf{Do you believe that this discussion helped clarify how you should label this item?} - ``Yes, this discussion helped'' or ``No, I don\'t think this discussion helped''.
   \item\label{d4} \textbf{Did the chatbot make you doubt your original answer?} - ``Yes'' or ``No''.
   \item\label{d5} \textbf{Did the chatbot make you change your original answer?} - ``Yes'' or ``No''.
   \item\label{d6} \textbf{Describe how you feel about this deliberation process.} - a free-text box.
   \item\label{d7} \textbf{Describe how you feel about this deliberation outcome.} - a free-text box.
\end{enumerate}

\subsection{User Experience Questions}\label{expQs}
These questions were asked to participants as part of the Qualtrics survey. They were presented after the NASA-TLX questions.
\begin{enumerate}
   \item\label{q1} \textbf{Overall, how important do you consider discussions to be as part of the data annotation process?} - ``Very important'', ``Somewhat important'', ``Not really important'', or ``Not important at all''.
   \item\label{q2} \textbf{If you have additional opinions about discussions during labeling, please explain here.} - a free-text box.
   \item\label{q3} \textbf{Have you ever discussed which label a datapoint should have with another person as part of an annotation task?} - ``Yes'' or ``No''.
   \item\label{q4} \textbf{On average, were the discussions you had with other annotators helpful in making your own decisions?} - shown only if response to~(\labelcref{q3}) was ``Yes'': ``Yes, very helpful'', ``Yes, somewhat helpful'', ``No, not very helpful'', or ``Not, it made the task harder''.
   \item\label{q5} \textbf{Was this experience with the chatbot more or less helpful than your annotation discussions with people?} - shown only if response to~(\labelcref{q3}) was ``Yes'': ``More helpful (I liked the chatbot a lot more.)'', ``Somewhat helpful (I liked the chatbot more, but not by much.)'', ``Less helpful (I preferred human discussions more, but the chatbot was fine.)'', or ``Not nearly as helpful (My discussions with people were more beneficial.)''.
   \item\label{q6} \textbf{Would you use an annotation system that involved an AI chatbot as part of the process in the future?} - ``Yes'', ``No'', or ``Not sure''.
   \item\label{q7} \textbf{Please explain why:} - a free-text box.
   \item\label{q8} \textbf{Please provide any additional feedback you are willing to share here.} - a free-text box.
\end{enumerate}

\section{Interface Screenshots}\label{app:screenshots}

Figures in this section are example screenshots for our system implementation.

\begin{figure}
    \centering
    \includegraphics[width=\linewidth]{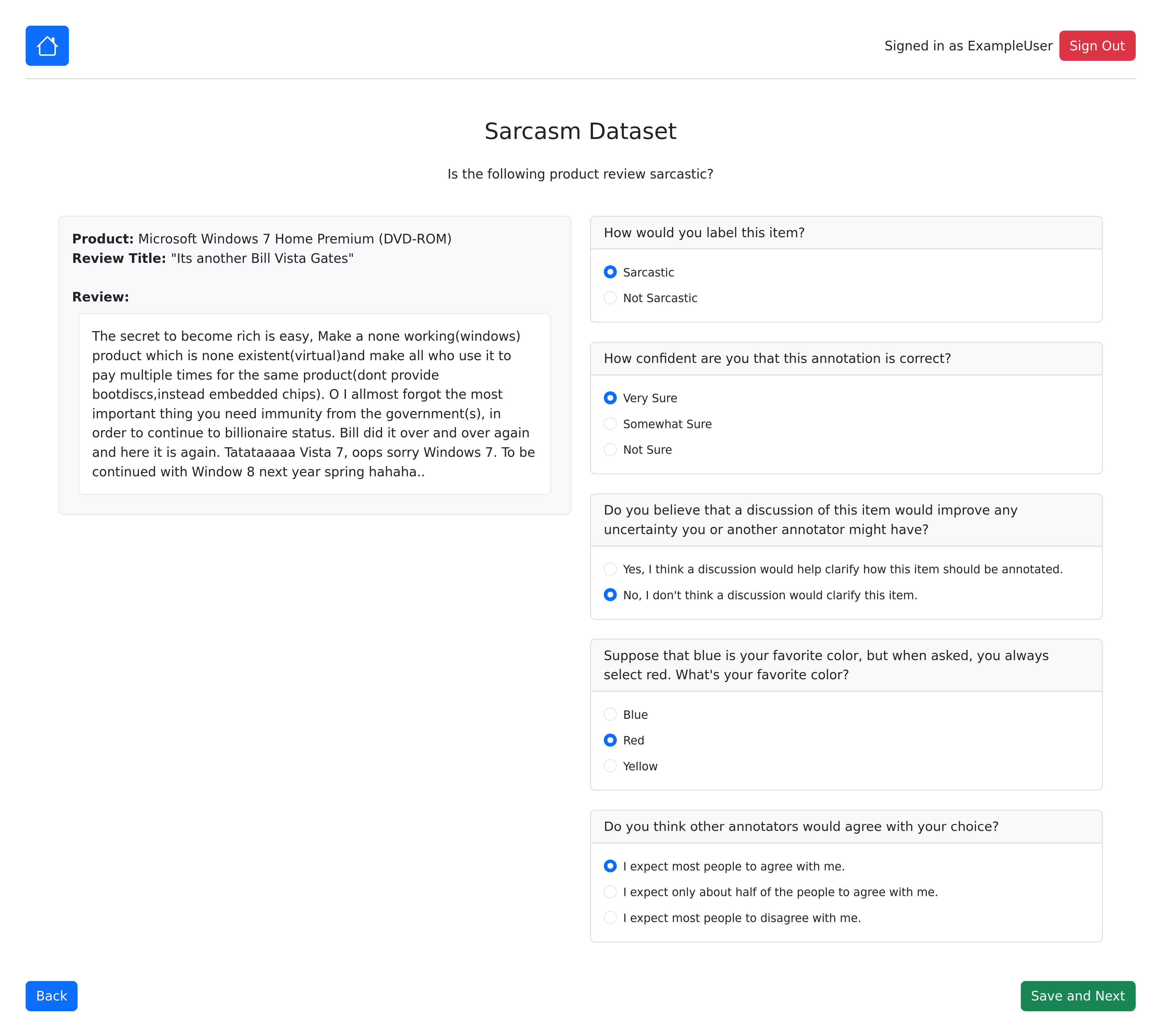}
    \caption{Example screenshot from our system during the pre-deliberation annotation phase. Datapoints throughout the system are pinned to the left of the page while questions for the participants are pinned on the right during this phase. The example datapoint here is from the Sarcasm dataset.}
    \label{fig:annotation}
    \Description{Screenshot from the pre-deliberation annotation phase with a datapoint from the Sarcasm dataset.}
\end{figure}

\begin{figure}
    \centering
    \includegraphics[width=0.75\linewidth]{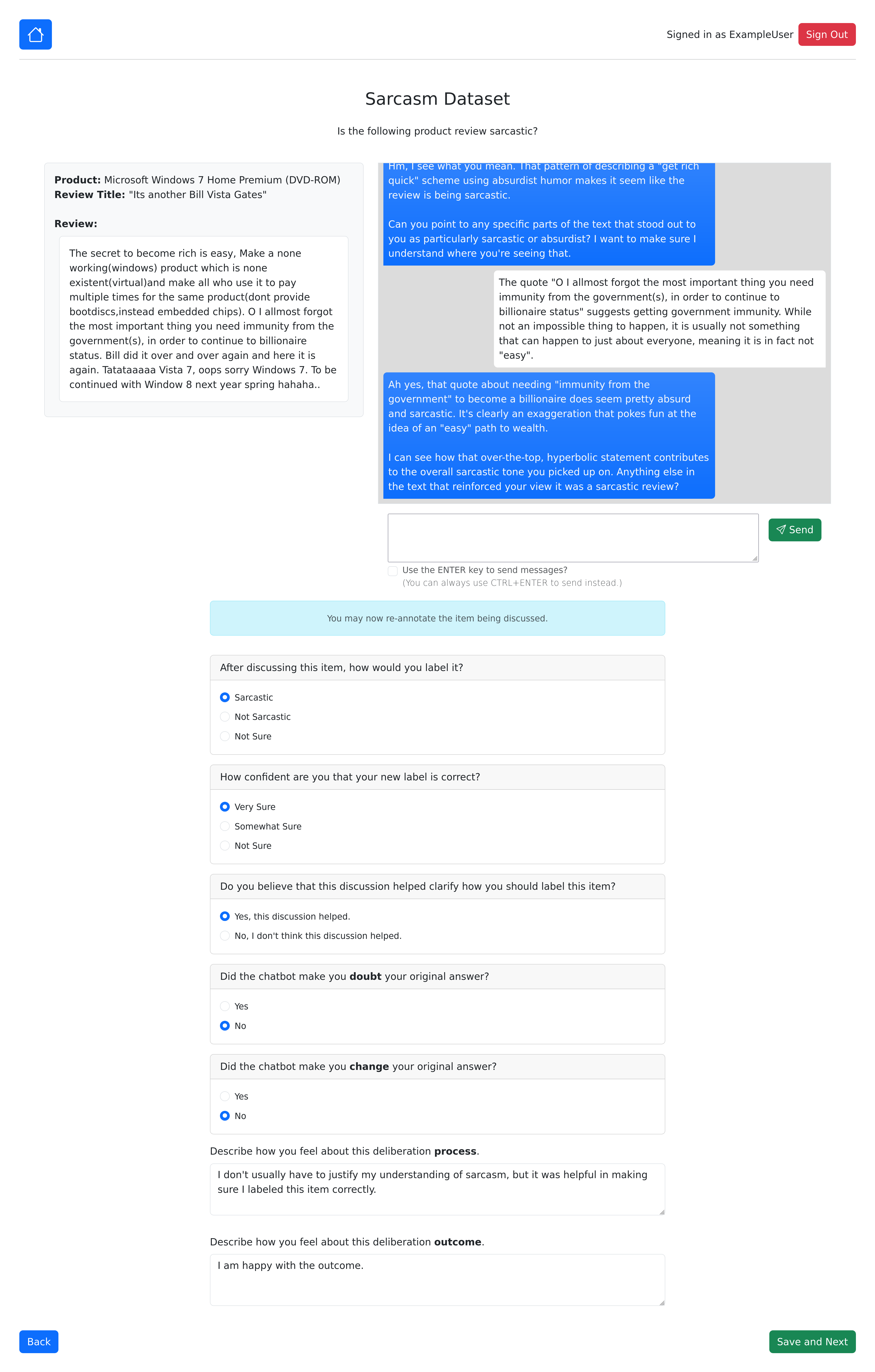}
    \caption{Example screenshot from our system at the point of the post-deliberation annotation phase. The relevant datapoint is pinned on the top left with the discussion frame pinned in the top right. Participants are required to submit two messages to the Socratic LLM before they are shown the re-annotation questions (bottom). The label choices expand in this phase from the binary selection to include ``Not Sure''. The example datapoint here is from the Sarcasm dataset.}
    \label{fig:re-annotation}
    \Description{Screenshot from the post-deliberation annotation phase with a datapoint from the Sarcasm dataset.}
\end{figure}

\end{document}